\documentclass[]{spie}

\usepackage{amsmath,amsfonts,amssymb,graphicx,xspace}
\usepackage[colorlinks=true, allcolors=blue]{hyperref}

\def\m87{M87$^*$\xspace}
\def\sgra{Sgr~A$^*$\xspace}

\def\lsim{\mathrel{\raise.3ex\hbox{$<$\kern-.75em\lower1ex\hbox{$\sim$}}}}
\def\gsim{\mathrel{\raise.3ex\hbox{$>$\kern-.75em\lower1ex\hbox{$\sim$}}}}

\title{The Black Hole Explorer: Motivation and Vision}

\author[1,2]{Michael D. Johnson}
    \affil[1]{Center for Astrophysics $|$ Harvard \& Smithsonian, 60 Garden Street, Cambridge, MA 02138, USA}
    \affil[2]{Black Hole Initiative at Harvard University, 20 Garden Street, Cambridge, MA 02138, USA}
\author[3,4,2]{Kazunori~Akiyama}
    \affil[3]{Haystack Observatory, Massachusetts Institute of Technology, Westford, MA 01886, USA}
    \affil[4]{Mizusawa VLBI Observatory, National Astronomical Observatory of Japan, Iwate 023-0861, Japan}
\author[1]{Rebecca~Baturin}
\author[5]{Bryan~Bilyeu}
    \affil[5]{MIT Lincoln Laboratory, Lexington, MA 02421}
\author[1,2]{Lindy~Blackburn}
\author[5]{Don~Boroson}
\author[6]{Alejandro~C\'ardenas-Avenda\~no}
    \affil[6]{Princeton Gravity Initiative, Princeton University, Princeton, New Jersey 08544, USA}
\author[6]{Andrew~Chael}
\author[7,8,9]{Chi-kwan~Chan}
    \affil[7]{Steward Observatory and Department of Astronomy, University of Arizona, 933 N. Cherry Ave., Tucson, AZ 85721, USA}
    \affil[8]{Data Science Institute, University of Arizona, 1230 N. Cherry Ave., Tucson, AZ~85721, USA}
    \affil[9]{Program in Applied Mathematics, University of Arizona, 617 N. Santa Rita, Tucson, AZ~85721, USA}
\author[1,2]{Dominic~Chang}
\author[1]{Peter~Cheimets}
\author[7]{Cathy~Chou}
\author[1,2]{Sheperd~S.~Doeleman}
\author[10,11]{Joseph~Farah}
    \affil[10]{Las Cumbres Observatory, 6740 Cortona Drive, Suite 102, Goleta, CA 93117-5575, USA}
    \affil[11]{Department of Physics, University of California, Santa Barbara, CA 93106-9530, USA}
\author[2,12,13]{Peter~Galison}
    \affil[12]{Department of History of Science, Harvard University, Cambridge, MA 02138, USA}
    \affil[13]{Department of Physics, Harvard University, Cambridge, MA 02138, USA}
\author[14]{Ronald~Gamble}
    \affil[14]{NASA Goddard Space Flight Center, Greenbelt, MD 20771, USA}
\author[15]{Charles~F.~Gammie}
    \affil[15]{Departments of Astronomy and of Physics, University of Illinois, Urbana, IL 61801, USA}
\author[6]{Zachary~Gelles}
\author[16]{José~L.~Gómez}
    \affil[16]{Instituto de Astrof\'{\i}sica de Andaluc\'{\i}a-CSIC, Glorieta de la Astronom\'{\i}a s/n, E-18008 Granada, Spain}
\author[17]{Samuel~E.~Gralla}
    \affil[17]{Department of Physics, University of Arizona, Tucson, AZ 85719, USA}
\author[1]{Paul~Grimes}  
\author[18,19]{Leonid~I.~Gurvits}
    \affil[18]{Joint Institute for VLBI ERIC, 7991\,PD Dwingeloo, The Netherlands}
    \affil[19]{Faculty of Aerospace Engineering, Delft University of Technology, 2629\,HS Delft, The~Netherlands}    
\author[20,21]{Shahar~Hadar}
    \affil[20]{Department of Mathematics and Physics, University of Haifa at Oranim, Kiryat Tivon 3600600, Israel}
    \affil[21]{Haifa Research Center for Theoretical Physics and Astrophysics, University of Haifa, Haifa 3498838, Israel}
\author[1]{Kari~Haworth}
\author[4,22]{Kazuhiro~Hada}
    \affil[22]{Department of Astronomical Science, The Graduate University for Advanced Studies (SOKENDAI), 2-21-1 Osawa, Mitaka, Tokyo 181-8588, Japan}
\author[3]{Michael~H.~Hecht}
\author[4,22,23]{Mareki~Honma}
    \affil[23]{Department of Astronomy, Graduate School of Science, The University of Tokyo, 7-3-1 Hongo, Bunkyo-ku, Tokyo 113-0033, Japan}
\author[1]{Janice~Houston}
\author[19,24]{Ben~Hudson}
    \affil[24]{KISPE Space Systems Limited, Farnborough, United Kingdom}
\author[1,2]{Sara~Issaoun}
\author[25]{He~Jia}
    \affil[25]{Department of Astrophysical Sciences, Princeton, NJ 08540, USA}
\author[26]{Svetlana~Jorstad}
    \affil[26]{Institute for Astrophysical Research, Boston University, 725 Commonwealth Ave., Boston, MA 02215, USA}
\author[3]{Jens~Kauffmann}
\author[27]{Yuri~Y.~Kovalev}
    \affil[27]{Max-Planck-Institut f\"ur Radioastronomie, Auf dem H\"ugel 69, D-53121 Bonn, Germany}
\author[14]{Peter~Kurczynski}
\author[14]{Robert~Lafon}
\author[28]{Alexandru~Lupsasca}
    \affil[28]{Department of Physics \& Astronomy, Vanderbilt University, Nashville, TN 37212, USA}
\author[29]{Robert~Lehmensiek}
    \affil[29]{National Radio Astronomy Observatory, Charlottesville, VA 22903, USA} 
\author[30,31]{Chung-Pei~Ma}
    \affil[30]{Department of Physics, University of California, Berkeley, CA 94720, USA}
    \affil[31]{Department of Astronomy, University of California, Berkeley, CA 94720, USA}
\author[7]{Daniel~P.~Marrone}
\author[26]{Alan~P.~Marscher}
\author[1]{Gary~J.~Melnick}
\author[1,2]{Ramesh~Narayan}
\author[32]{Kotaro~Niinuma}
    \affil[32]{Graduate School of Sciences and Technology for Innovation, Yamaguchi University, Yamaguchi 753-8512, Japan}
\author[14]{Scott~C.~Noble}
\author[14]{Eric~J.~Palmer}
\author[1,2]{Daniel~C.~M.~Palumbo}
\author[3]{Lenny~Paritsky}
\author[14]{Eliad~Peretz}
\author[1,2]{Dominic~Pesce}
\author[1,2]{Alexander~Plavin}
\author[6,25]{Eliot~Quataert}
\author[1,2]{Hannah~Rana}
\author[1,2]{Angelo~Ricarte}
\author[1,2]{Freek~Roelofs}
\author[5]{Katia~Shtyrkova}
\author[33]{Laura~C.~Sinclair}
    \affil[33]{National Institute of Standards and Technology, Boulder, Colorado 80305}
\author[14]{Jeffrey~Small}
\author[29]{Sridharan~Tirupati~Kumara}
\author[1]{Ranjani~Srinivasan}
\author[2,13,34]{Andrew~Strominger}
    \affil[34]{Center for the Fundamental Laws of Nature, Harvard University, Cambridge, MA, USA}
\author[1,2]{Paul~Tiede}
\author[1]{Edward~Tong}  
\author[5]{Jade~Wang}
\author[1]{Jonathan~Weintroub}
\author[27]{Maciek~Wielgus}
\author[6,35]{George~Wong}
    \affil[35]{School of Natural Sciences, Institute for Advanced Study, 1 Einstein Drive, Princeton, NJ 08540, USA}
\author[1,2]{Xinyue~Alice~Zhang}

\authorinfo{Corresponding Author: Michael D.\ Johnson (\url{mjohnson@cfa.harvard.edu})}

\begin{document} 
\maketitle
\newpage

\begin{abstract}
We present the Black Hole Explorer (BHEX), a mission that will produce the sharpest images in the history of astronomy by extending submillimeter Very-Long-Baseline Interferometry (VLBI) to space. BHEX will discover and measure the bright and narrow ``photon ring'' that is predicted to exist in images of black holes, produced from light that has orbited the black hole before escaping. This discovery will expose universal features of a black hole's spacetime that are distinct from the complex astrophysics of the emitting plasma, allowing the first direct measurements of a supermassive black hole’s spin. In addition to studying the properties of the nearby supermassive black holes \m87 and \sgra, BHEX will measure the properties of dozens of additional supermassive black holes, providing crucial insights into the processes that drive their creation and growth. BHEX will also connect these supermassive black holes to their relativistic jets, elucidating the power source for the brightest and most efficient engines in the universe. BHEX will address fundamental open questions in the physics and astrophysics of black holes that cannot be answered without submillimeter space VLBI. The mission is enabled by recent technological breakthroughs, including the development of ultra-high-speed downlink using laser communications, and it leverages billions of dollars of existing ground infrastructure. We present the motivation for BHEX, its science goals and associated requirements, and the pathway to launch within the next decade.

\end{abstract}


\section{Introduction}

Black holes are central to questions of stellar evolution, galaxy formation and mergers, astrophysical jets, and the nature of spacetime. They are prodigious energy sources, liberating gravitational energy to power the brightest objects in the Universe \cite{Schmidt_1963,Lynden-Bell_1969,Shakura_Sunyaev,Novikov_Thorne,Yuan_Narayan_2014}. 
Supermassive black holes are ubiquitous in galactic cores \cite{Richstone_1998,Magorrian_1998,Fabian2012,Kormendy_2013}, and their spin can launch and power relativistic jets that extend over thousands of parsecs, ultimately shaping evolution on galactic scales \cite{Blandford_Znajek,Tchekhovskoy_2011,Blandford_2019}.
Yet, the vast influence of black holes originates from their defining property on the smallest scale: a gravitational singularity enshrouded within an event horizon.

Over the past decade, our ability to directly study event-horizon-scale physics using images has been revolutionized by the Event Horizon Telescope (EHT).
The EHT is an Earth-spanning very-long-baseline interferometry (VLBI) array that has produced images at ${\sim}$230\,GHz (corresponding to a wavelength of $\lambda=1.3\,{\rm mm}$) \cite{EHTC_M87_I,EHTC_M87_II,EHTC_M87_III,EHTC_M87_IV,EHTC_M87_V,EHTC_M87_VI,EHTC_M87_VII,EHTC_M87_VIII,EHTC_M87_IX,EHTC_SgrA_I,EHTC_SgrA_II,EHTC_SgrA_III,EHTC_SgrA_IV,EHTC_SgrA_V,EHTC_SgrA_VI,EHTC_SgrA_VII,EHTC_SgrA_VIII}.
The longest baselines of the EHT have a length of $d \approx 11{,}000\,{\rm km}$, with a corresponding angular resolution $\lambda/d \approx 20\,\mu{\rm as}$.
This resolution is only a few times larger than the angular gravitational scales $\theta_{\rm g} \equiv GM/(c^2 D)$ of its two primary targets: {\m87} ($\theta_{\rm g} \approx 4\,\mu{\rm as}$), the $M \approx 6 \times 10^9 M_\odot$ black hole at the center of the elliptical galaxy M87, located $D \approx 53$ million light years away, and {\sgra} ($\theta_{\rm g} \approx 5\,\mu{\rm as}$), the $M \approx 4 \times 10^6 M_\odot$ black hole at the center of the Milky Way, $D \approx 27{,}000$ light years away.
Previous VLBI studies of \m87 and \sgra---even those with space VLBI \cite{Johnson_2021,Kim_2023}---could not access these scales because they were performed at lower frequencies, where synchrotron self-absorption (for both targets) and interstellar scattering (for \sgra) entirely obscure horizon-scale features.
The combination of the sharp angular resolution and high observing frequencies with the EHT revealed a dark central region (the ``apparent shadow'' \cite{Falcke2000}) in each source, with a diameter of about $10\theta_{\rm g}$.

\begin{figure}
    \centering
    \includegraphics[width=0.83\textwidth]{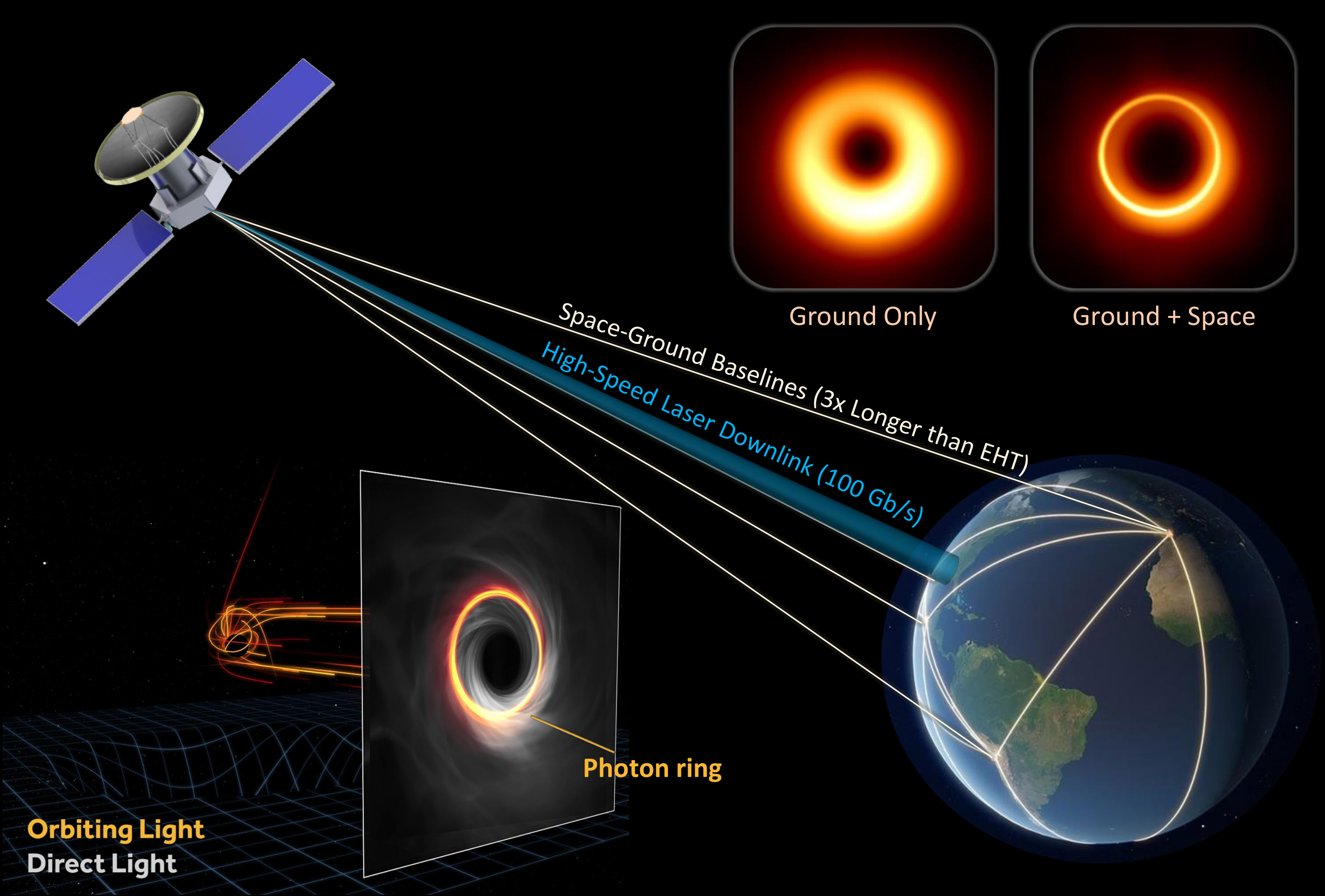}
    \caption{
    {\bf The BHEX mission concept.}
    Black hole images display distinctive, universal features such as a sharp ``photon ring'' that is produced from light that has orbited the black hole before escaping.
    By extending the Earth-spanning EHT into space, BHEX will be the first mission to make precise measurements of this striking, untested prediction from general relativity, enabling the first direct measurement of a supermassive black hole's spin.
    }
    \label{fig:BHEX}
\end{figure}

\begin{figure}
    \centering
    \includegraphics[width=\textwidth]{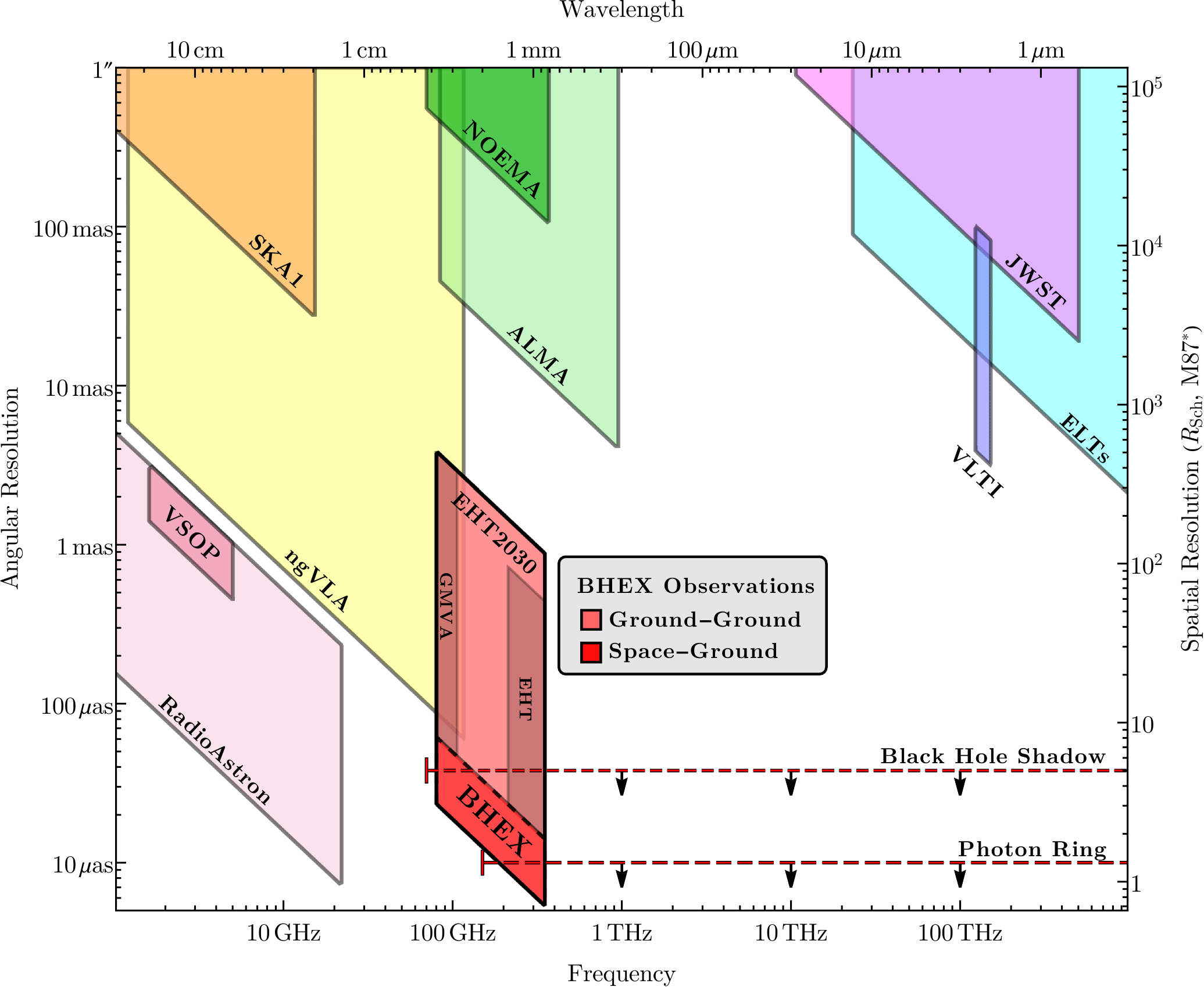}
    \caption{
    {\bf BHEX: Opening a new window on the Cosmos.}
    BHEX will deliver the sharpest images in the history of astronomy, accessing structures that no other facility can reach.
    Ground-based VLBI has steadily improved its capabilities over the past decades, with facilities spanning the globe. At the highest frequencies, the EHT has resolved the apparent shadows of the \m87 and \sgra black holes and is being significantly enhanced through programs including the next-generation EHT (ngEHT\cite{ngEHT_refarray,ngEHT_KSG}) to include new telescopes and to span frequencies from 80-360\,GHz with simultaneous multi-band observations. However, ground-based VLBI cannot produce images at frequencies above ${\sim}400\,$GHz because of severe atmospheric turbulence and absorption \cite{Pesce_2024}.
    Past space-VLBI missions (VSOP\cite{Hirabayashi_1998} and RadioAstron\cite{Kardashev_2013}) observed only at frequencies below ${\sim}22\,$GHz, where relativistic features such as the black hole shadow and photon ring cannot be seen because of synchrotron self-absorption.
    By working with ground millimeter/submillimeter facilities, BHEX will be the first mission to overcome both limitations, thereby revealing a black hole's photon ring.
    Figure adapted from Refs.~[\citenum{Selina_2018,ngEHT_KSG}].
    }
    \label{fig:BHEX_parameter_space}
\end{figure}

The EHT images have had a profound influence across the astronomy and physics communities, and they prompted a worldwide explosion of interest in black holes from the general public.
The EHT image of \m87 features prominently in the Astro2020 Decadal Survey \cite{astro2020}, which noted the important role that such observations could play in two of its three organizing science themes---``Cosmic Ecosystems'' and ``New Messengers and New Physics''---particularly in pursuit of answers to fundamental questions about accretion, jets, galaxy evolution, and strong-field gravity.
At EHT resolution, we are beginning to see the structures controlling inflow and outflow around the black hole, but are limited in what we can discern about spacetime itself or the mechanism through which black holes anchor, launch, and power their relativistic jets \cite{Narayan_Quataert_2023}.
For example, while the EHT images give the masses of \m87 and \sgra to an accuracy of ${\sim}10\%$, they do not directly constrain either black hole's spin.
Finer details in the images of black holes, just beyond the reach of an Earth-bound VLBI array, reflect crisp relativistic features and would enable truly fundamental discoveries. 

Recently, we have shown that general relativity predicts a universal feature of black hole images: a sharp ring of light formed by photons that escape after orbiting near the event horizon, known as the ``photon ring'' \cite{Johnson_2020,Gralla_Lupsasca_2020a}. 
This photon ring arises from the family of unstable spherical photon orbits near a Kerr black hole \cite{Bardeen_1973,Teo_2003,Yang_2012}.  
The properties of the black hole, including its spin, are elegantly encoded in the photon ring, as is the origin of black hole feedback power \cite{Johnson_2020,Himwich_2020,Gralla_Lupsasca_2020a,PWP,Chael_2023}. 
For ground-ground VLBI baselines, the photon ring is blended with the more weakly lensed emission \cite{Tiede_2022}, but it can be isolated using space-ground VLBI baselines where the signal is dominated by the sharpest image features \cite{Johnson_2020,Gelles_2021,AART}.
Moreover, it can be studied using a sparse interferometric array because of a telltale ``ringing'' pattern in interferometric space \cite{Johnson_2020,Gralla_2020,Gralla_Lupsasca_2020b,GLM2020,Paugnat2022,Vincent_2022,Cardenas_2023,Cardenas_2024,Jia_2024}.

Motivated by the promise of such measurements, we have developed a concept for a mission that would extend the EHT to space: the {\it Black Hole Explorer} (BHEX; see Figures~\ref{fig:BHEX} and \ref{fig:BHEX_parameter_space}). 
This work has included contributions from over 100 scientists and engineers, focused workshops on science and engineering \cite{Kurczynski_2022,BHEX_Peretz_2024}, white papers summarizing the potential capabilities of space VLBI and exploring a variety of mission concepts \cite{Roelofs_2019,Astro2020_Origins,Lazio_2020,Gurvits_2021,Gurvits_2022,Hudson2023}, detailed studies on crucial technologies \cite{Wang_2023}, and extensive international contributions that include a burgeoning consortium of scientists and engineers from the Japanese space community \cite{BHEX_Akiyama_2024}. 
BHEX will operate as a ``hybrid observatory'' \cite{Mather_2024,BHEX_Issaoun_2024}, combining mission-critical infrastructure on the ground and in space.
In particular, these efforts to develop BHEX are catalyzed by intensive parallel efforts to expand the ground-based capabilities of the EHT through higher-frequency observations \cite{Crew_2023,EHT_345}, major upgrades and expansion of the array \cite{Backes_2016,Asada_2017,Romero_2020,Kauffmann_2023,ngEHT_refarray}, frequency phase transfer through simultaneous multi-band observations \cite{Rioja_2015,Issaoun_2023,Rioja_2023,Jiang_2023}, and enhancements of the most sensitive ground facilities such as the ALMA Observatory \cite{Carpenter_2023,Navarrini_2023}. 

In this paper, we describe our vision for BHEX and the technology and instrumentation that makes it possible to launch as a NASA Explorers mission within the next decade.
We begin by summarizing the science goals (Section~\ref{sec:Science_Goals}) and their associated requirements (Section~\ref{sec:Requirements}).
We then describe the instrument and its associated heritage (Section~\ref{sec:Instrument}) and the concept of operations (Section~\ref{sec:ConOps}).
We conclude with a summary of the mission status and pathway to launch (Section~\ref{sec:Summary}).  
A series of associated white papers describe the BHEX instrument \cite{BHEX_Marrone_2024}, its associated subsystems \cite{BHEX_Tong_2024,BHEX_Sridharan_2024,BHEX_Rana_2024,BHEX_Wang_2024,BHEX_Srinivasan_2024,BHEX_Tomio_2024}, the planned coordination with ground networks \cite{BHEX_Issaoun_2024}, as well as several of the primary science drivers for the mission \cite{BHEX_Lupsasca_2024,BHEX_Galison_2024,BHEX_Kawashima_2024}.

\section{Science Goals}
\label{sec:Science_Goals}

\begin{figure}
    \centering
    \includegraphics[width=\textwidth]{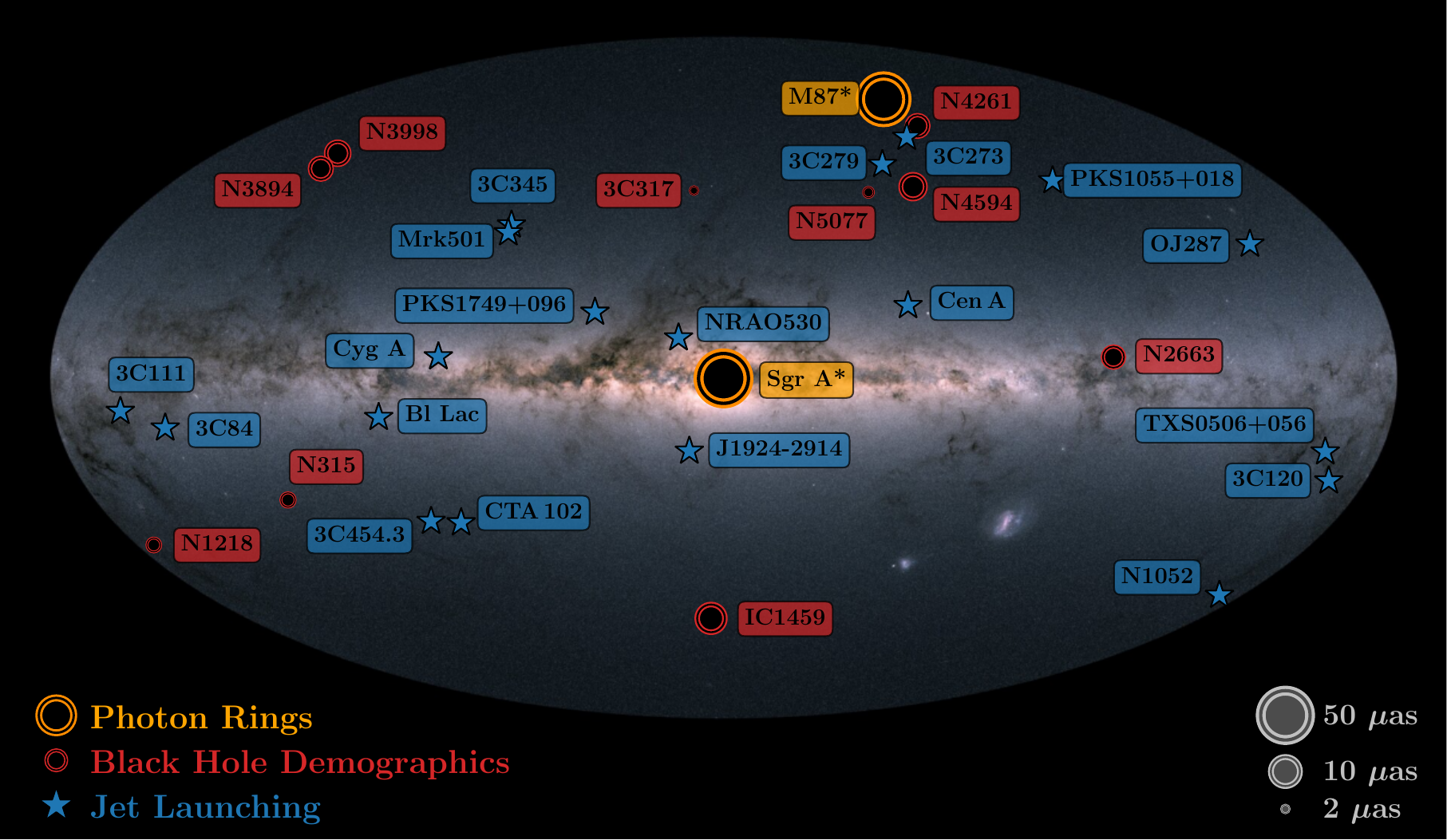}
    \caption{
    {\bf Potential BHEX targets.}
    Horizon-scale targets are shown as rings, scaled (logarithmically) by the expected angular size of the black hole shadow.
    The primary targets for photon ring studies and direct measurements of spin, \sgra and \m87, are shown in orange.
    The remaining horizon-scale targets, shown in red, include a variety of sources in low accretion states, sampling both radio-loud and radio-quiet sources \cite{Zhang_2024}.
    In addition, BHEX will study a wide variety of AGN with prominent jets, including potential binary supermassive black holes (e.g., OJ287), sources associated with neutrino emission (TXS0506${+}$056), and bright gamma-ray blazars with superluminal features (e.g., 3C279, CTA102).
    Many of these sources have already been studied with the EHT \cite{Kim_2020,Janssen_2021,Issaoun_2022,Jorstad_2023,Paraschos_2024}.
    Background image: ESA/Gaia/DPAC.
    }
    \label{fig:BHEX_Targets}
\end{figure}

We now summarize the primary science goals for BHEX.
These goals address some of the most pressing open questions in black hole science across three broad themes.
In Section~\ref{sec:Photon_Ring}, we describe how BHEX will reveal fundamental properties of supermassive black holes through studies of the photon rings in \m87 and \sgra.
In Section~\ref{sec:Jets}, we describe how BHEX will determine how supermassive black holes launch and accelerate relativistic jets through measurements of the jet collimation profile, polarization structure, and dynamics in a sample of active galactic nuclei.
Finally, in Section~\ref{sec:Demographics}, we show how BHEX will probe the growth of supermassive black holes by measuring the horizon-scale properties of a population of supermassive black holes in low accretion states across a variety of galaxy morphologies.
Figure~\ref{fig:BHEX_Targets} summarizes potential targets for BHEX across these three science goals.
For discussion of additional science opportunities with BHEX, including a potential single-dish mode, see Ref.~[\citenum{BHEX_Akiyama_2024}]. 

\subsection{Precision Measurements of Black Holes: the Photon Ring}
\label{sec:Photon_Ring}

\begin{figure}[t]
    \centering
    \includegraphics[width=\textwidth]{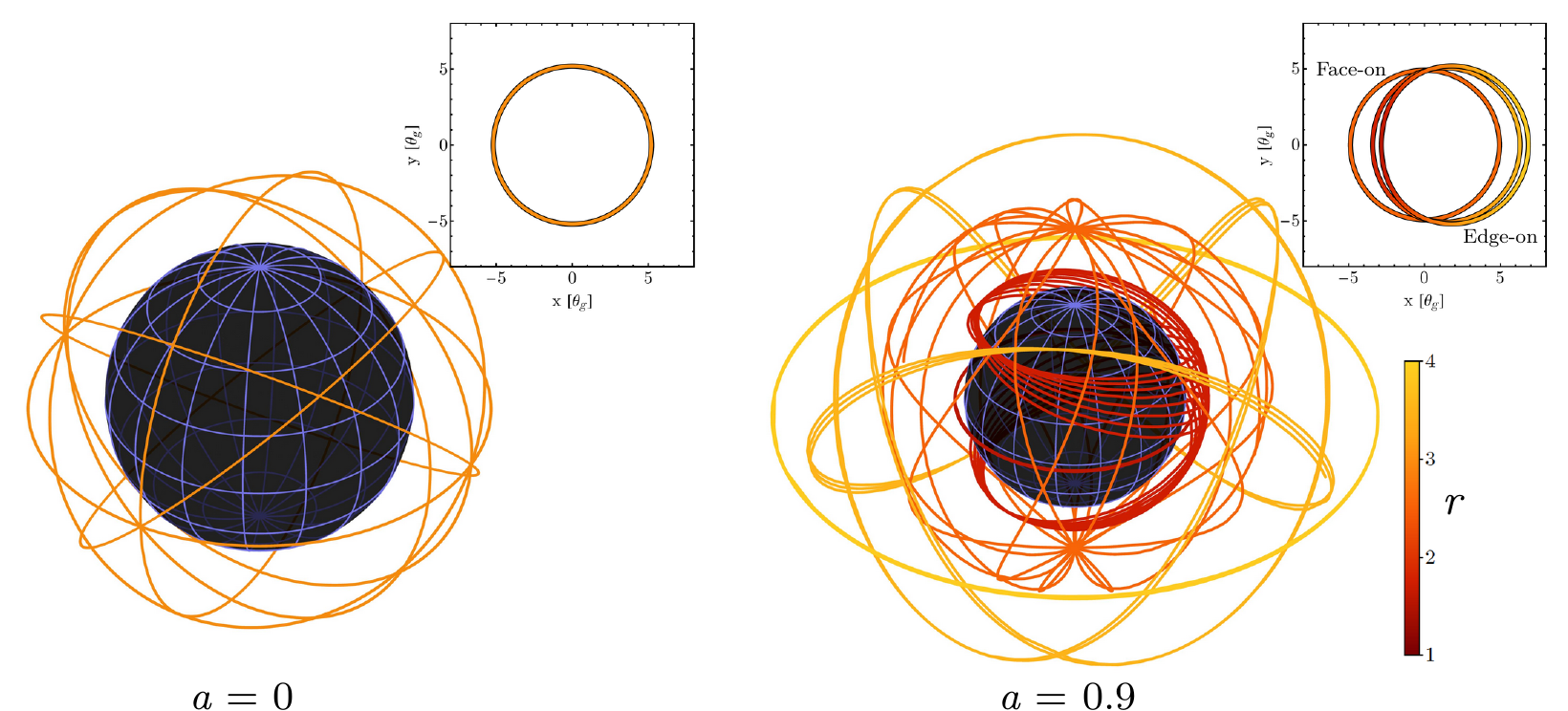}
    \caption{    
    {\bf The photon ring.}
    A black hole's photon ring arises from the family of unstable photon orbits near a black hole, which are bound on spherical surfaces (in Boyer-Lindquist coordinates) \cite{Bardeen_1973,Teo_2003}.
    The left panel shows orbits for a non-rotating black hole ($a=0$; the event horizon is at $r_{+}=2 GM/c^2$), which lie at a fixed radius $r\equiv 3 GM/c^2$ (defining the photon sphere) and give rise to an asymptotic photon ring that is circular with a radius $\sqrt{27}\theta_{\rm g}$ (shown as the inset).
    The right panel shows orbits for a rotating black hole ($a=0.9$; the event horizon is at $r_{+}=1+\sqrt{1-a^2}\approx 1.4 GM/c^2$), which span a range of radii (defining the photon shell) and give rise to an asymptotic photon ring that is both smaller and deformed, with a size and shape that are sensitive to the spin and viewing inclination relative to the spin axis \cite{Teo_2003,Johannsen_2010,Johnson_2020}.
    Remarkably, the image of the photon shell maps photon orbits at varying emission radius to varying image angle.
    A \href{https://youtu.be/79CF68h89GU}{video} showing these photon orbits is available online.
    }
\label{fig:Photon_Ring}
\end{figure}

Photons passing near a black hole can fall in or escape to infinity.
Between those two possibilities are unstable orbits that lie on spherical surfaces \cite{Bardeen_1973,Teo_2003}.
For a non-rotating (Schwarzschild) black hole, these orbits have a common radius (defining the ``photon sphere'').
For a spinning (Kerr) black hole, these orbits lie within a range of radii (defining the ``photon shell''). 

\begin{figure}[t]
    \centering
    \includegraphics[width=0.25\textwidth]{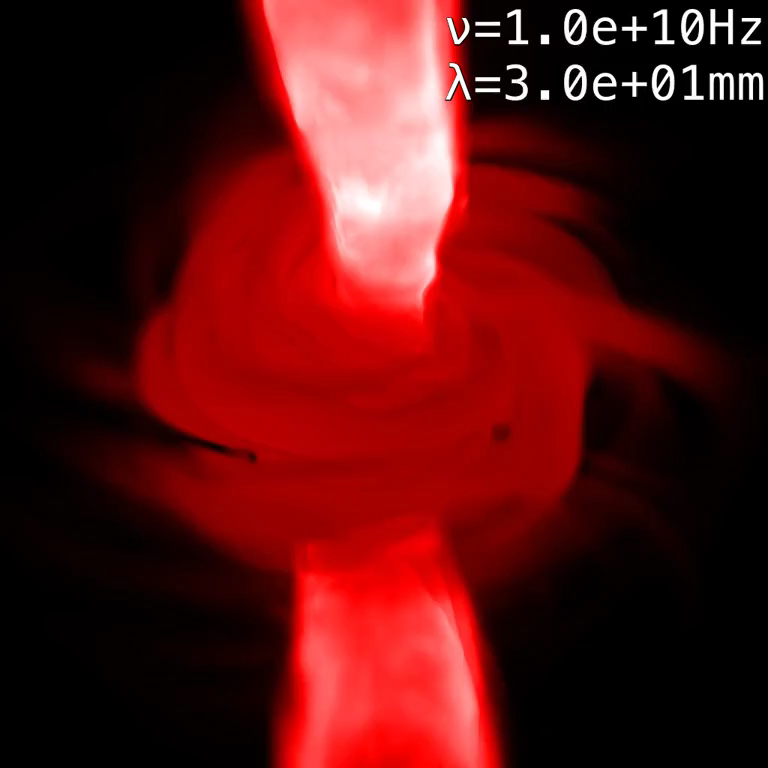}%
    \includegraphics[width=0.25\textwidth]{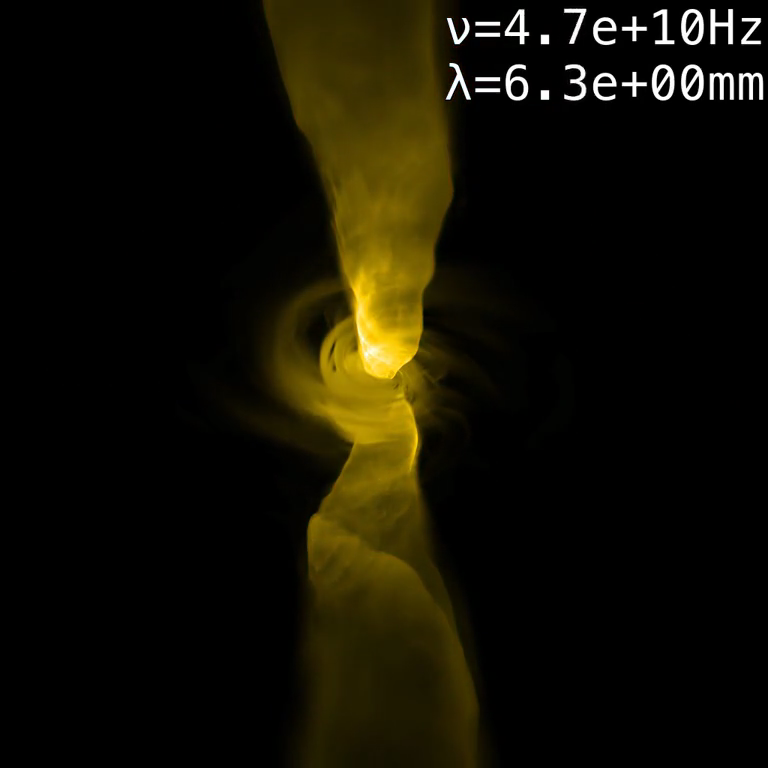}%
    \includegraphics[width=0.25\textwidth]{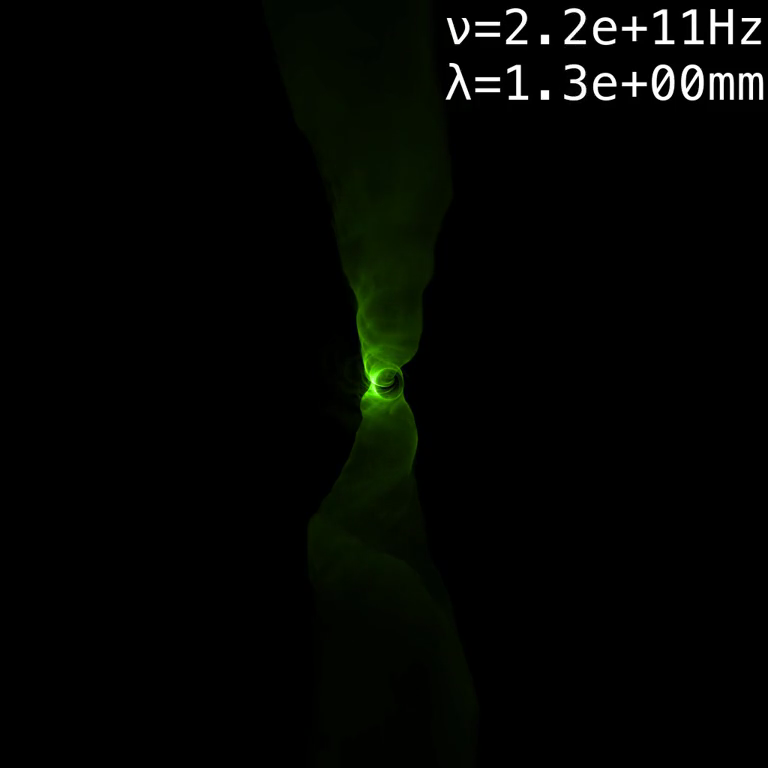}%
    \includegraphics[width=0.25\textwidth]{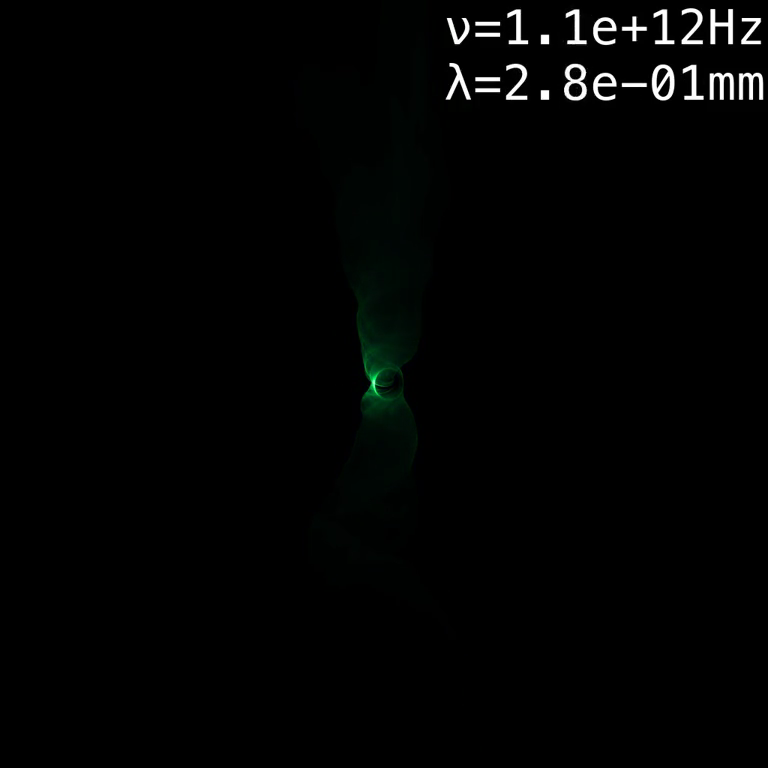}
    \caption{
    {\bf Model of \sgra shown in multiple wavelengths} \cite{Chan_2013}.
    Two panels on the left: For low frequencies (${<}\,100\,$GHz), the accretion flow is optically thick and blocks the photon ring.
    Two panels on the right: For higher frequencies (${>}\,200\,$GHz), the accretion flow becomes optically thin and the photon ring is visible.
    A \href{https://youtube.com/shorts/Qrly-3_FNpc}{video} showing the opacity with changing frequency is available online.
    }
\label{fig:sgra_nu}
\end{figure}

This shell of unstable spherical photon orbits produces a striking image feature, produced from light that has orbited the black hole before escaping: the photon ring (see Figures~\ref{fig:BHEX} and \ref{fig:Photon_Ring}) \cite{Johnson_2020,BHEX_Lupsasca_2024}.
Because the photon ring is a result of gravitational lensing, it is generically seen on simulated images \cite{Gralla_2019,Johnson_2020,Vincent2022}, including those from state-of-the-art general relativistic magnetohydrodynamic (GRMHD) simulations \cite{Gammie2003,Porth2019,EHTC_M87_V,EHTC_SgrA_V}, whenever the emission is optically thin~(see Figure~\ref{fig:sgra_nu}).
In addition, the photon ring is a {\it universal} signature---the surrounding matter has negligible effects on the trajectory of light.
Thus, resolved measurements of the photon ring present an extraordinary opportunity for detailed studies of a black hole's spacetime.

\begin{figure}
    \centering
    \includegraphics[width=0.7\textwidth]{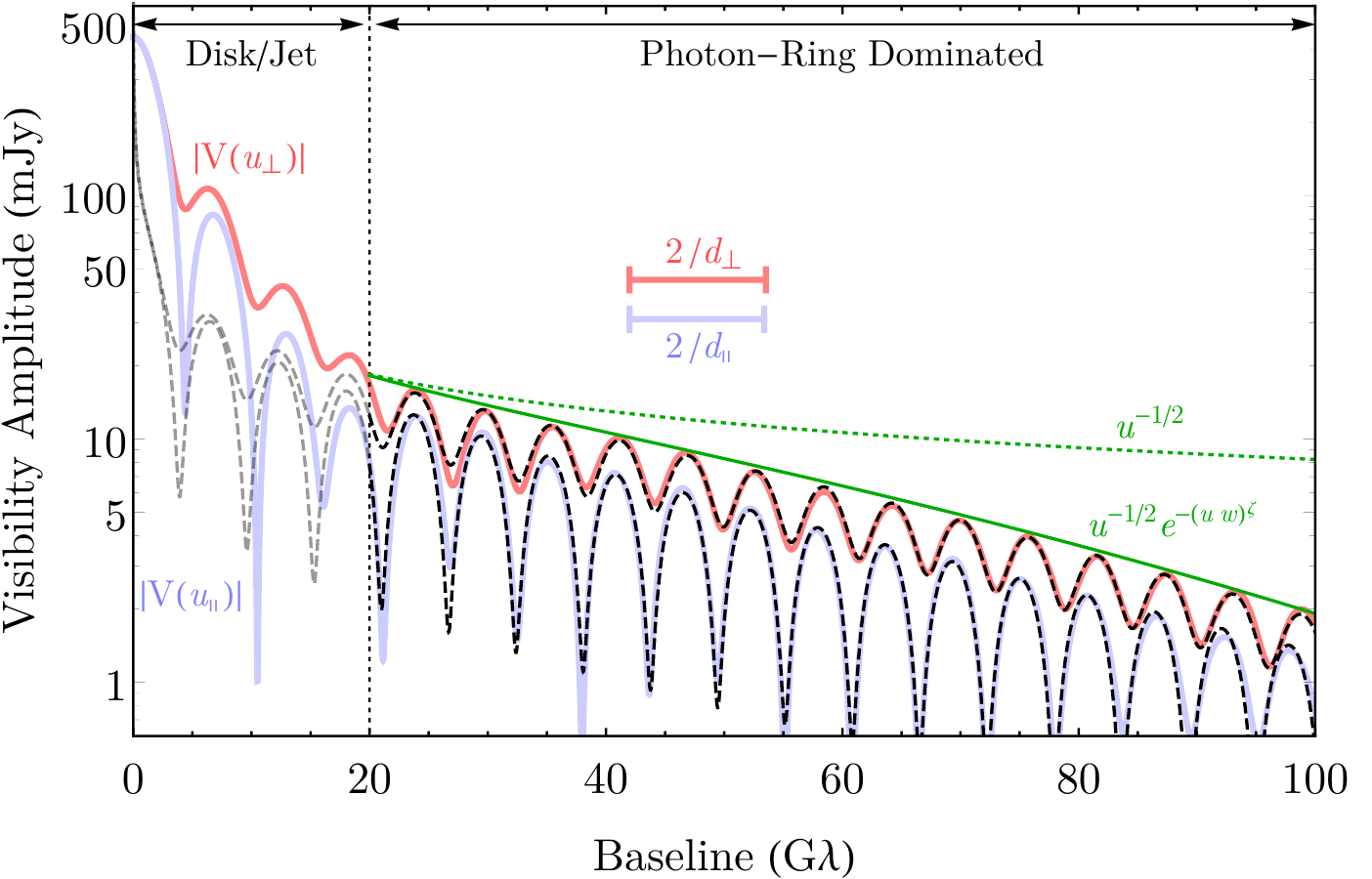}
    \caption{
    {\bf Universal interferometric signature of the photon ring.}
    The black hole photon ring produces a strong interferometric signature in the radio visibility of the source.
    Here, we show the periodic signal produced in a time-averaged GRMHD-simulated image of \m87.
    This clear ringing dominates the signal on long baselines to space and displays a periodicity that encodes the angle-dependent diameter of the photon ring \cite{Gralla_2020,Gralla_Lupsasca_2020b,GLM2020}.
    Measuring this periodicity on the space baselines targeted with BHEX will provide a clear photon ring detection and measurement of its shape, which encodes information about the black hole mass and spin \cite{BHEX_Lupsasca_2024}.
    Reproduced from Ref.~[\citenum{Johnson_2020}].
    }
    \label{fig:uv_plot}
\end{figure}

\begin{figure}
    \centering
    \includegraphics[width=\textwidth]{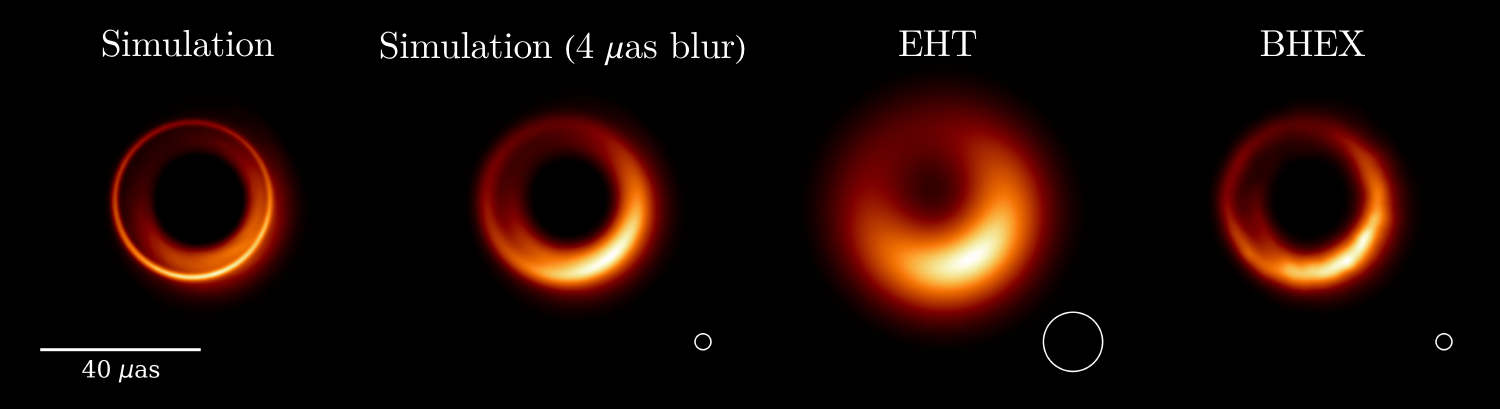}
    \caption{
    {\bf Black hole imaging with BHEX.}
    The left panel shows a time-averaged image of \m87 from a GRMHD simulation, which displays prominent relativistic features such as the photon ring and inner shadow \cite{Chael_2021}.
    The left-center panel shows the simulated image convolved with a $4\,\mu{\rm as}$ Gaussian beam.
    The right-center panel shows an image reconstruction at the approximate resolution of the current EHT ($15\,\mu{\rm as}$).
    The right panel shows a reconstruction with BHEX at 240\,GHz, averaging 30 simulated observations over a 3-month window (${\approx}240 GM/c^3$) \cite{Comrade}.
    These observations used a ground array with 10 existing ground telescopes (but excluding the ALMA array).
    The improved angular resolution with BHEX accesses the sharp features, which are blended together at the resolution of the EHT.
    }
    \label{fig:BHEX_Imaging_Example}
\end{figure}

The most technically demanding goal of BHEX is to discover the predicted photon rings in \m87 and \sgra, with measurements that are precise enough to directly measure the black hole spins.
In particular, BHEX can provide a firm, model-independent detection of the photon ring by measuring its distinctive interferometric signature for the time-averaged signal on long baselines: a damped periodic signal as a function of baseline length (see Figure.~\ref{fig:uv_plot}) \cite{Johnson_2020}.
Because the dimensionless baseline length $u=d/\lambda$ depends on both the physical baseline length $d$ and the observing wavelength, $\lambda$, a single ground-space baseline can see the periodic oscillation in $u$ as long as the stations in the baseline instantaneously capture a wide enough range of wavelengths.
Thus, BHEX has the ability to uncover this prediction of general relativity through its recently discovered imprint that is only accessible on space-VLBI baselines \cite{Johnson_2020,Himwich_2020,Gralla_2020,Wielgus2021,Palumbo_2022,Tiede_2022,Cardenas_2023}, as well as through direct imaging (see Figure~\ref{fig:BHEX_Imaging_Example}).

\begin{figure}[t]
    \centering
    \includegraphics[width=\textwidth]{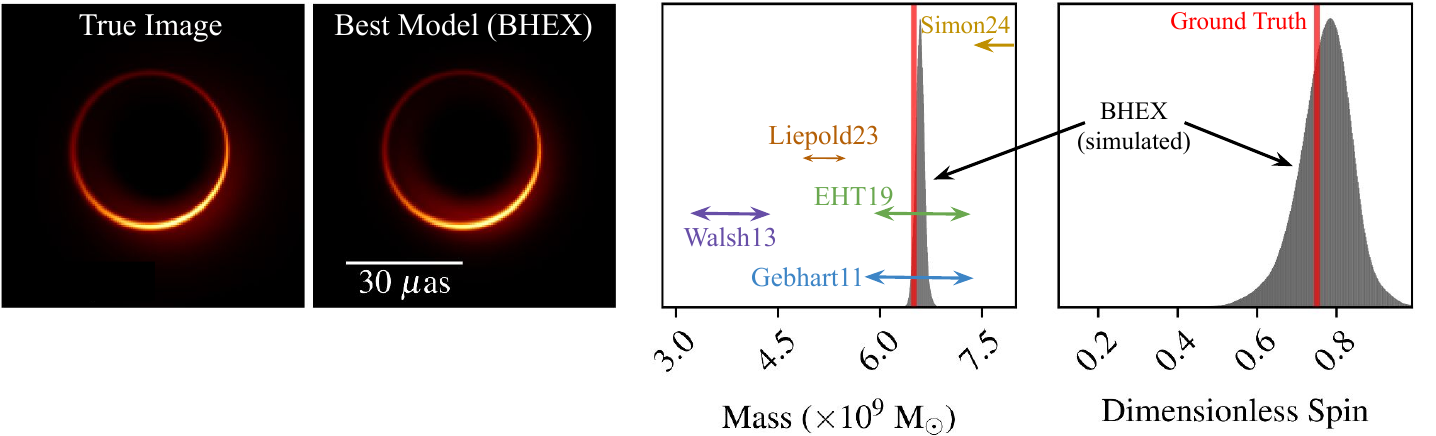}
    \caption{    
    {\bf Measuring black hole spin with BHEX.}
    These panels show the results of model fits to a simulated image of \m87 (left) using software that simultaneously fits for the black hole spacetime and the plasma parameters in a Bayesian framework \cite{Palumbo_2022b}.
    The marginalized posteriors on the black hole mass and dimensionless spin are shown in the rightmost two panels (the red vertical lines show the correct GRMHD values for this test).
    The maximum likelihood fitted model is also shown for comparison.
    These fits indicate that BHEX can make measurements of the black hole mass that are sufficiently precise and accurate to resolve tension among even modern measurements of the \m87 mass across a variety of methods \cite{Gebhardt_2011,Walsh_2013,Liepold_2023,Simon_2024,EHTC_M87_VI}.
    More importantly, BHEX can constrain the black hole spin to ${\sim}$10\% accuracy, even accounting for systematic uncertainties from the complex astrophysical environment. 
    }
\label{fig:Spin_Measurement}
\end{figure}

BHEX measurements of the photon ring would provide unambiguous measurements of black hole spin (see Figure~\ref{fig:Spin_Measurement}).
Specifically, a spinning black hole has a non-circular photon ring, with a degree of asymmetry that increases with the magnitude of the black hole's spin \cite{Takahashi_2004,Farah_2020,Paugnat2022}.
Thus, precise measurements of the photon ring offer an opportunity to make this fundamental measurement of the spin of \m87 and to test whether the black hole's spin is indeed the energy source for its powerful jet \cite{Blandford_Znajek,PaperV,Chael_2023}.
We note that there are no previous measurements of spin in supermassive black holes that are not strongly dependent on astrophysical assumptions.
One of the most relied-upon methods currently uses models of accretion disks and their predicted x-ray reflection spectra to match against observed fluorescent emission lines; the precise shape and width of these emission lines depends on the depth to which the accretion disk's inner edge lies, which in turn depends on the black hole spin and the assumption that the reflected emission truncates at this inner edge \cite{Bambi2021,Reynolds2021}.
For additional discussion of the details and scientific opportunities related to photon ring measurements for BHEX, see Refs.~[\citenum{BHEX_Lupsasca_2024,BHEX_Galison_2024}].

The photon ring also provides an avenue to explore exotic alternatives to Kerr black holes.
At the resolution of the EHT, these alternatives can effectively mimic the appearance of a Kerr black hole \cite{Vincent_2016,Mizuno_2018,Olivares_2020,Vincent_2021,FPWP}.
However, they can produce strikingly different photon rings at scales accessible to BHEX.
For instance, boson stars---macroscopic quantum objects with no hard surface, no horizon, and no singularity---may possess no photon shell and thus exhibit no photon ring in the image, depending on their discreet angular momentum and scalar field compactness \cite{Cunha2016}.
By contrast, images of wormholes can produce multiple photon rings, composed of light bent around photon shells on both sides of the wormhole \cite{Wielgus_2020}.
For additional discussion of how extensions of the EHT can be used to constrain fundamental physics, see Ref.~[\citenum{FPWP}].


\subsection{The Origin of Relativistic Jets from Supermassive Black Holes}
\label{sec:Jets}

One of the most remarkable byproducts of black hole accretion is the production of relativistic jets.
These jets are among the most energetic phenomena in our universe, emitting radiation throughout the entire electromagnetic spectrum, from radio wavelengths to the $\gamma$-ray regime, and even accelerating particles to the highest measured energies \cite{Blandford_2019}.
High-resolution VLBI observations and numerical simulations have advanced our understanding of the accretion processes and the ensuing jet formation.
Observations with the EHT have enabled us to probe the innermost regions of the accretion flow and the base of the jet in unprecedented detail, revealing intricate structures and magnetic field configurations crucial for jet launching \cite{Chael_2023}.
These findings support the hypothesis that the interplay between the black hole spin and the magnetic field topology near the event horizon plays a pivotal role in jet formation.
Moreover, GRMHD simulations have demonstrated that the efficiency of jet production is strongly dependent on the black hole spin parameter and the magnetic flux accumulated in the vicinity of the black hole \cite{Blandford_Znajek}.
Despite these advancements, the exact processes governing the conversion of accreted mass and energy into collimated, relativistic outflows remain largely unknown.

Among the various types of active galactic nuclei (AGN), blazars are particularly notable for their highly energetic jets pointed almost directly toward Earth.
Mapping the innermost regions of blazar jets to understand how they are launched, collimated, and accelerated, as well as the role played by magnetic fields in these processes, requires state-of-the-art VLBI observations with the highest possible angular resolution.
Space-VLBI observations at centimeter wavelengths with RadioAstron have enabled the mapping of blazar jets with angular resolutions on the order of tens of microarcseconds.
These observations have revealed how the jet in the powerful radiogalaxy 3C84 is collimated\cite{Giovannini_2018}, the presence of helical magnetic fields launching powerful relativistic jets in BL Lac\cite{Gomez_2016}, and how the development of Kelvin-Helmholtz instabilities leads to the formation of filaments in the jet of 3C279\cite{Fuentes_2023}, structures that remain hidden when observed with ground-only arrays.
Conversely, ground-based millimeter VLBI observations with the Event Horizon Telescope offer the highest possible angular resolution achievable from the ground, reaching 20\,$\mu$as.
These observations operate at wavelengths where the blazar sources are optically thin, allowing us to penetrate the deepest regions of blazars and study areas closest to the black hole, which remain hidden behind an opacity curtain at other wavelengths.
The EHT observations of the blazar 3C279 have revealed an unexpected twisted and bent jet structure close to the central black hole\cite{Kim_2020}, while observations of Centaurus A have detailed the jet collimation and its connection to the supermassive black hole\cite{Janssen_2021}, enhancing our understanding of jet formation and acceleration mechanisms.
These results (see also Refs.~[\citenum{Issaoun_2022,Jorstad_2023,Paraschos_2024}]) highlight the transformative potential of the EHT in advancing our knowledge of relativistic jets in AGN.
The next natural step to obtain the sharpest view of blazar jets as close as possible to the central black hole is to perform space-VLBI observations at millimeter and submillimeter wavelengths with BHEX, capable of achieving angular resolutions of just a few microarcseconds.

Another key aspect to understand the nature of blazar jets is to decipher where and how the high energy emission observed from these objects is produced.
Multi-wavelength observations have shown that flares in $\gamma$-rays are usually preceded by a rapid rotation of the optical polarization angle at optical and millimeter wavelengths, which provides evidence for the existence of a helical magnetic field in the innermost jet regions, where the plasma is accelerated and collimated \cite{Marscher_2008,Marscher_2010,Jorstad_2010}.
Correlations among the different spectral bands have also allowed to locate the site of the $\gamma$-ray emission parsecs away from the central engine in OJ287, AO 0235+164, 3C120, and CTA102 \cite{Agudo_2011,Agudo_2011b,Casadio_2015, Casadio_2015b,Lico_2022}. 

In recent years, there has been a growing association between flaring activity in blazars and neutrino events.
High-energy neutrino detections by telescopes like IceCube have been temporally linked to blazar flares, as exemplified by the case of TXS 0506+056 \cite{IceCube_2018,IceCube_2018b}.
Further studies have shown a statistically significant association between neutrino sources and the bright cores of blazars, suggesting that these jets are likely sites of neutrino production \cite{Plavin_2020}.
The correlation between a sharp gamma-ray and optical flare with the 8 December 2022 neutrino detected toward 0735+178 adds to the evidence supporting this connection \cite{Acharyya_2023}.

Figure.~\ref{fig:BHEX_Targets} illustrates some of the most promising AGN targets for the BHEX mission.
These targets include potential supermassive binary black hole systems, neutrino candidates, nearby AGN where we can study jet formation and dynamics as close as possible to the central engine, and some of the brightest $\gamma$-ray sources in the sky.

\subsection{Supermassive Black Hole Demographics and Growth}
\label{sec:Demographics}

Finally, while BHEX will only be able to probe the photon rings for the most massive and nearby SMBHs (primarily \m87 and \sgra), it will be able to spatially resolve the horizon-scale emission for black holes that are substantially less massive or more distant, providing images comparable in quality to the current EHT images of \m87 and \sgra.
The ring-like structures seen by the EHT towards \m87 and \sgra are expected to be ubiquitous for accreting black holes observed in the optically thin regime, and the angular diameter of the emission ring $\theta_{\rm r}$ is proportional to the black hole's mass-to-distance ratio: $\theta_{\rm r} \approx 10 M/D$ \cite{PaperV}.

\begin{figure}[t]
    \centering
    \includegraphics[width=0.9\textwidth]{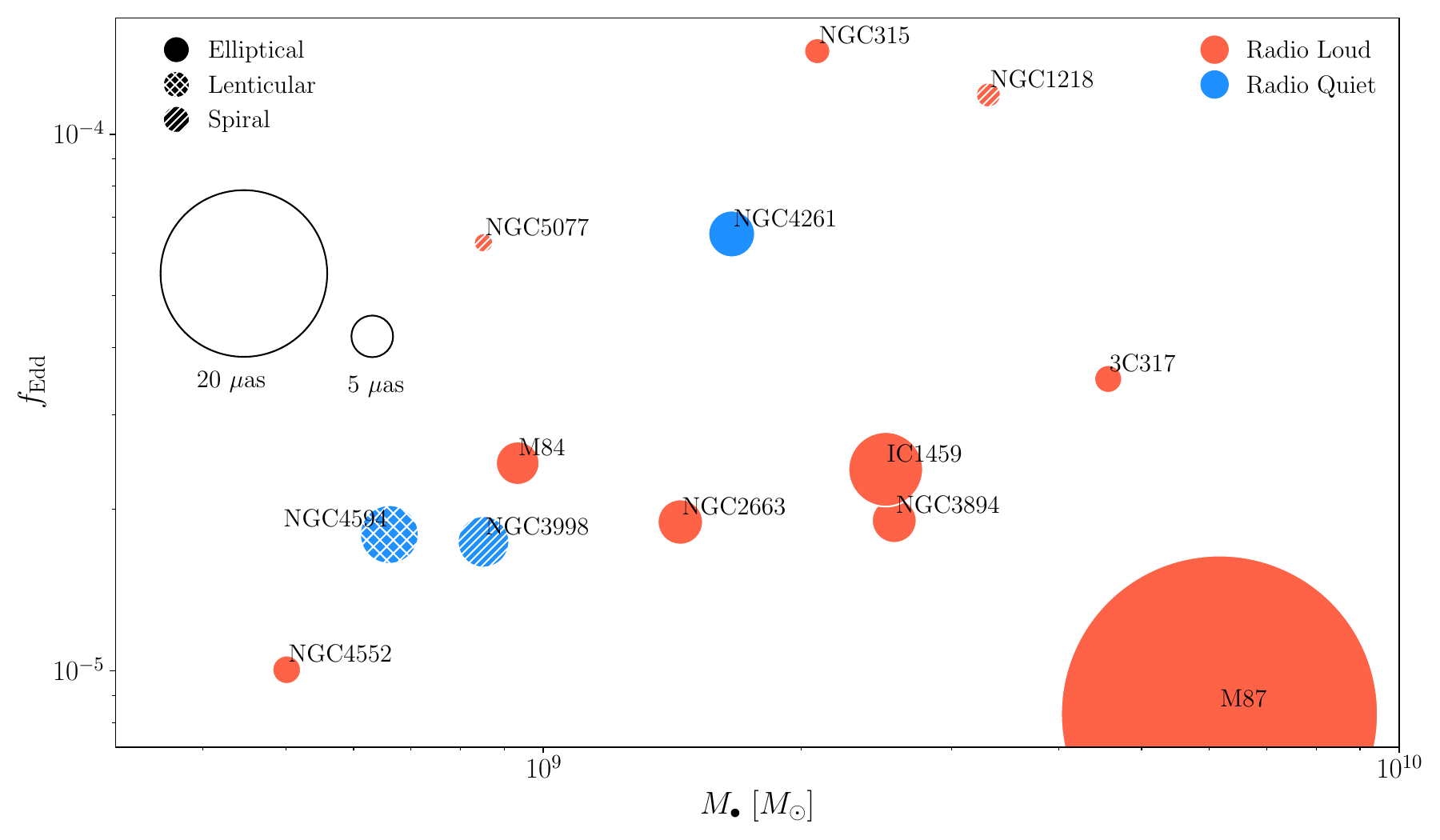}
    \caption{    
    {\bf Candidate BHEX targets for horizon-scale studies.}
    The increased resolution provided by BHEX will enable access to a population of black holes in a variety of different environments.
    This figure shows estimated Eddington ratio as a function of black hole mass for \m87 and 12 additional sources.
    The symbol color encodes the radio loudness of each source (based on the 5\,GHz to X-ray flux ratio \cite{Wang_FP_2024}), the shading encodes the morphology of the host galaxy, and the size scales linearly with the estimated photon ring diameter.
    Adapted from Ref.~[\citenum{Zhang_2024}].
    }
\label{fig:Demographics}
\end{figure}

The angular resolution of BHEX will in principle be sufficient to spatially resolve dozens of additional SMBH shadows in the nearby ($z \lesssim 0.1$) Universe \cite{Pesce_2021}.
Unlike most AGN samples, which are biased towards (rarer) high-Eddington ratio objects, BHEX targets for horizon-scale studies will be predominantly those with low Eddington ratios, more representative of the typical accretion states of supermassive black holes.
Figure.~\ref{fig:Demographics} shows the estimated Eddington ratios and black hole masses for a subset of candidate BHEX targets from Ref.~[\citenum{Zhang_2024}], selected from the ongoing Event Horizon and Environs (ETHER) survey \cite{Ramakrishnan_2023,Hernandez_2024}.  Eddington ratios are estimated based on the 230\,GHz flux density; black hole masses are typically derived from dynamical modeling.

By expanding the sample of horizon-resolved black holes beyond merely \sgra and \m87, BHEX will enable demographic studies in poorly explored regions in parameter space.  These dimensions include the following: 

\begin{itemize}
    \item \underline{Black Hole Mass}: By measuring source sizes and applying the techniques already developed and employed for \m87 \cite{EHTC_M87_VI} and \sgra \cite{EHTC_SgrA_IV,EHTC_SgrA_VI}, BHEX will substantially increase the number of SMBHs with precisely-measured masses.
    Interestingly, the sample of promising BHEX targets overlaps with the list of objects most likely to produce observable nHz gravitational waves \cite{Mingarelli_2017}.
    The large amplitude of the nHz gravitational wave background has motivated scrutiny of the high-mass end of the black hole mass function  \cite{Sato-Polito_2023,Izquierdo-Villalba_2024}.
    Direct mass measurements with BHEX will synergize with gravitational wave science to help resolve these newfound tensions.
    \item \underline{Eddington Ratio}: The study of Ref.~[\citenum{Zhang_2024}] predicts that most new targets should  have larger Eddington ratios than those inferred for either \m87 or \sgra ($\sim$$10^{-5}$ and $\sim$$10^{-7}$ respectively \cite{EHTC_M87_VIII,EHTC_SgrA_VIII}), up to $\sim$$10^{-4}$ \cite{Zhang_2024}.
    At these higher accretion rates, radiative cooling will become important \cite{Yoon_2020}, which is typically neglected for \sgra and \m87.
    \item \underline{Spin}: If the underlying population of black holes spans a range in spin, then we may observe spin-driven differences in accretion disk and jet activity.
    The Blandford-Znajek mechanism predicts that the jet power should be proportional to the square of the spin  \cite{Blandford_Znajek}, which can be directly tested using samples such as those presented in Figure.~\ref{fig:Demographics} that include both radio-loud and radio-quiet AGN.
    The distribution of SMBH spins in the local Universe should encode details of their growth histories, including spin-down from jets \cite{Narayan_2022,Ricarte_2023_spindown}, SMBH-SMBH mergers \cite{Berti&Volonteri2008,Ricarte_2023}, and the relative frequency of prograde versus retrograde accretion \cite{King_2008}.
    \item \underline{Inclination}: Our views of \m87 and \sgra are close to face-on \cite{Walker_2018,EHTC_SgrA_VIII}.
    The general population should naturally exhibit a uniform distribution of viewing angles.
    Polarized emission should evolve substantially as a function of inclination due to geometric effects and Faraday rotation \cite{Qiu_2023,EHTC_SgrA_VIII}.
    Polarized imaging of the expanded BHEX sample will enable inferences of the three-dimensional structure of accretion disks and jets. 
    \item \underline{Environment}: The most promising candidate targets for BHEX include elliptical galaxies, lenticulars, and even spirals (e.g., the Sombrero Galaxy, NGC~4594).
    Not all ellipticals are central cluster galaxies like M87, and at least one is at the center of a less massive galaxy group (NGC4261; \cite{Davis_1995}).
    Thus, studies with BHEX in combination with images on larger scales will connect the accretion and jet launching of black holes to their environment in a variety of intra- and inter-galactic environments. 
\end{itemize}

\section{Requirements}
\label{sec:Requirements}

\subsection{General Considerations for Space VLBI}

The most demanding requirements for BHEX are those related to detecting and measuring the photon rings in \m87 and \sgra.
This requires both that 1) the photon ring is visible on the sky and that 2) BHEX has sufficient angular resolution and sensitivity to measure the photon ring.  

\begin{figure}
    \centering
    \includegraphics[width=\textwidth]{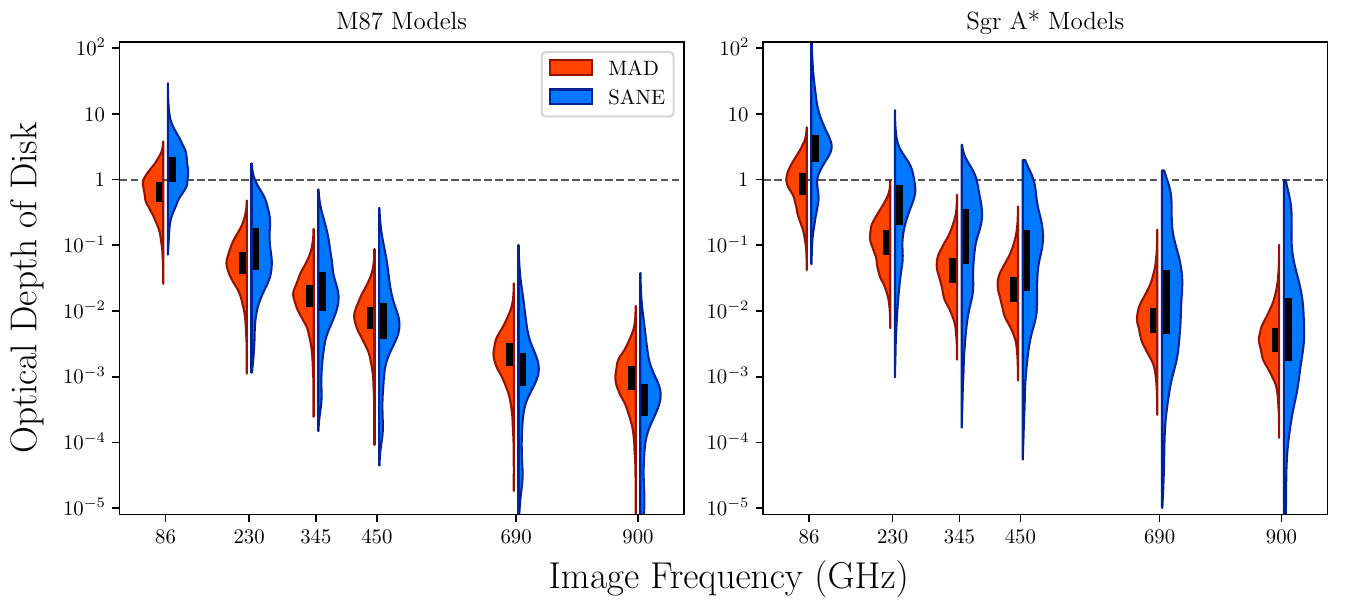}
    \caption{Image-averaged optical depth as a function of observing frequency over a suite of GRMHD models for \m87 (left) and \sgra (right).
    The models cover both MAD and SANE configurations, varying black hole spins, and a variety of electron temperature models.
    For \sgra, the images also include varying inclinations.
    Reproduced from Ref.~[\citenum{Wong_2024}].
    }
    \label{fig:Optical_Depth}
\end{figure}

To ensure that the photon ring is visible, BHEX must observe \m87 and \sgra at high enough frequencies that optical depth to synchrotron self-absorption does not hide the photon ring.
General-relativistic magnetohydrodynamic simulations have been extremely successful in producing models that are consistent with EHT observations of \m87 and \sgra and can generate images with angular resolution of $\mu{\rm as}$ or less, providing a firm basis to derive requirements for BHEX.
These simulations have been developed for decades and are highly constrained by EHT measurements \cite{EHTC_M87_V,EHTC_M87_VIII,EHTC_M87_IX,EHTC_SgrA_V,EHTC_SgrA_VIII}.
As shown in Figure~\ref{fig:Optical_Depth}, these simulations indicate that BHEX must observe above 200\,GHz to ensure that the optical depth is $\tau < 1$, allowing access to the photon rings in \m87 and \sgra.

In addition, radio observations are affected by lines of sight through the ionized interstellar medium, which has an index of refraction that varies with the local electron density, introducing scattering \cite{Rickett_1990,Narayan_1992}.
While scattering is negligible for most sources at millimeter and submillimeter wavelengths, the scattering toward \sgra is approximately 1000 times stronger than typical lines of sight \cite{Davies_1976}.
Scattering obscures the photon ring and reduces the flux density on long baselines \cite{Zhu_2019}.
Thus, while angular resolution requirements set a lower limit on the length of BHEX baselines, scattering sets an upper limit on the useful baseline length for a given observing frequency or, equivalently, a lower limit on the observing frequency for a given baseline: 
\begin{align}
	\nu\gsim280\,{\rm GHz} \left(\frac{b}{20{,}000\,{\rm km}}\right).
\end{align}
The scattering has been studied intensively at centimeter wavelengths \cite{Shen_2005,Bower_2006,Gwinn_2014}, giving rise to a detailed predictive model for BHEX \cite{Psaltis_2018,Johnson_2018}.
Because of ISM scattering, BHEX must observe above 300\,GHz to access the photon ring in \sgra. Because the scattering of \m87 is $1{,}000$ times weaker than that of \sgra, this requirement only affects \sgra. 

Finally, because the interferometric resolution of BHEX is determined by its separation from the ground-based stations, its orbit is tuned to see spatial scales where the photon ring dominates over smoother, more weakly lensed emission.
For \m87 and \sgra, this transition occurs at approximately $20\,{\rm G}\lambda$.

\subsection{Key Performance Metrics}

BHEX is primarily a mission for continuum VLBI, limited by angular resolution ($\lambda/D$) and sensitivity.
The angular resolution of a baseline depends only on the length $b$ of the baseline projected orthogonal to the line of sight and on the observing wavelength:
\begin{align}
    \theta &\approx 10\,\mu{\rm as} \times \left( \frac{\lambda}{1\,{\rm mm}} \right) \left( \frac{b}{20{,}000\,{\rm km}} \right)^{-1}\\
    &\approx \left( 8\,\mu{\rm as} \right) \times \left(\frac{\lambda}{1\,{\rm mm}} \right) \left( \frac{\text{Altitude}}{20{,}000\,{\rm km}} \right)^{-1}
\end{align}

The sensitivity of a radio telescope is commonly quantified by its system equivalent flux density (SEFD) \cite{TMS}:
\begin{align}
    {\rm SEFD} &= \frac{2k_{\rm B} T_{\rm sys}^\ast}{\eta_{\rm A} A}.
\end{align}
In this expression, $k_{\rm B}$ is the Bolzmann constant, $T_{\rm sys}^\ast$ is the effective system temperature, $\eta_{\rm A}$ is the aperture efficiency of the telescope, and $A$ is the geometric area of the dish.
The aperture efficiency depends on a number of factors including the reflector design and losses from surface deformations with RMS surface errors of $\epsilon$ (Ruze losses):
\begin{align}
    \eta_{\rm A,Ruze} = e^{-\left(\frac{4\pi \epsilon}{\lambda} \right)^2}
\end{align}
For example, to ensure that $\eta_{\rm A,Ruze} > 0.75$ for the highest observing frequency of BHEX (320\,GHz) requires $\epsilon < 40\,\mu{\rm m}$.

Because BHEX does not have any contribution from atmospheric emission or absorption, $T_{\rm sys}^\ast$ is approximately equal to the receiver noise temperature, $T_{\rm R}$.
Putting in characteristic values, we obtain 
\begin{align}
    {\rm SEFD} &\approx 20{,}000\,{\rm Jy} \times \left( \frac{T_{\rm R}}{50\,{\rm K}} \right) \left( \frac{D}{3.5\,{\rm m}} \right)^{-2} \left( \frac{\eta_{\rm A}}{0.7} \right)^{-1}\\
    &\approx 10{,}000\,{\rm Jy} \times \left( \frac{\nu}{100\,{\rm GHz}} \right) \left( \frac{D}{3.5\,{\rm m}} \right)^{-2}  \left( \frac{\eta_{\rm R}}{0.2} \right)^{-1} \left( \frac{\eta_{\rm A}}{0.7} \right)^{-1}.
\end{align}
In the final expression, we have introduced $\eta_{\rm R} \equiv T_{\rm R}/T_{\rm Q}$, which gives the performance of a receiver relative to its quantum limit $T_{\rm Q} \equiv h \nu/k_{\rm B} \approx 4.8\,{\rm K} \times \left( \frac{\nu}{100\,{\rm GHz}} \right)$ \cite{Caves_1982,Kerr_1997}.
To compute the SEFD for dual-sideband (DSB) receivers, the receiver noise temperature must be doubled to account for noise in both sidebands.

For continuum VLBI observations, which will be used for all the primary BHEX science goals, the most important quantity is the ground-space baseline sensitivity (expressed as the RMS thermal noise, $\sigma_{\rm G-S}$).
This quantity depends on the properties of both the ground and space telescopes:
\begin{align}
    \sigma_{\rm G-S} &= \frac{1}{\eta_{\rm Q}} \sqrt{ \frac{{\rm SEFD}_{\rm G}\, {\rm SEFD}_{\rm S}}{2\, \Delta \nu\, \Delta t}}.
\end{align}
Here, $\Delta \nu$ is the averaged bandwidth (single polarization), $\Delta t$ is the (coherent) integration time, and $\eta_{\rm Q} \leq 1$ is a factor that accounts for losses in coarse digitization of the electric field (for BHEX baselines, $\eta_{\rm Q}=0.75$; see Section~\ref{sec:DBE}).
We note that the baseline length does not affect the baseline sensitivity.
For the original EHT observations of \m87, the median SEFD at $\nu = 230\,$GHz across all sites was $5{,}000$\,Jy, with the best performance of 74\,Jy (a phased array of 37 12-m ALMA dishes).
Several other EHT sites (e.g., NOEMA and LMT) also achieve ${\rm SEFD} < 1{,}000\,{\rm Jy}$ in typical weather conditions for a source at moderately high elevation. 

Putting in characteristic values for strong baselines, we obtain
\begin{align}
    \sigma_{\rm G-S} &\approx 5\,{\rm mJy} \times \left( \frac{{\rm SEFD}_{\rm G}}{1{,}000\,{\rm Jy}} \right)^{1/2} \left( \frac{{\rm SEFD}_{\rm S}}{20{,}000\,{\rm Jy}} \right)^{1/2} \left( \frac{\Delta \nu}{8\,{\rm GHz}} \right)^{-1/2} \left( \frac{\Delta t}{100\,{\rm s}} \right)^{-1/2}.
\end{align}
All elements of the system combine to determine this baseline sensitivity, resulting in a rich trade space.  

Frequency phase transfer between the secondary ($\nu_2$) and primary ($\nu_1$) receivers for BHEX will require reaching a signal-to-noise ratio of approximately ${\rm S/N} \gsim \nu_1/\nu_2$ for the signal of the secondary receiver in the coherence time associated with the primary receiver.
For instance, if the coherence time at 320\,GHz is 10~seconds, frequency phase transfer from 80 to 320\,GHz would require reaching a signal-to-noise ratio at 80\,GHz of ${\rm S/N} \gsim 4$ in 10~seconds of integration.
The signal could then be coherently integrated in both bands for longer times (typically over a full ${\sim}10$-minute scan).
For additional details on frequency phase transfer, including results for a power-law spectrum of phase fluctuations, see Ref.~[\citenum{Pesce_2024}].

\subsubsection{Requirements for Photon Ring Detections}

The precise expected signal from a black hole depends on a multitude of details about the black hole system, many of which are poorly constrained.
However, in the optically thin regime (which is appropriate for observations of \m87 and \sgra at submillimeter wavelengths), the signal on long baselines takes a universal form that is determined solely by basic considerations of the gravitational lensing \cite{Johnson_2020}.
Specifically, ${\sim}20\%$ of the flux density comes from the black hole's ``photon ring,'' and this flux falls as $u^{-3/2}$ on long baselines as the photon ring is increasingly resolved.
Here, $u$ is the baseline length in units of the observing wavelength, with a corresponding angular resolution (or fringe spacing) $\theta = 1/u$. 

Because the flux density of the compact components of the sources such as \m87 and \sgra are known, we can compute the expected photon ring signal on a long baseline.
For instance, for \m87,
\begin{align}
    |V| &\approx 30\,{\rm mJy} \times \left( \frac{u}{10\,{\rm G}\lambda} \right)^{-3/2}\\
    &\approx 0.3\,{\rm mJy} \times \left( \frac{\theta}{1\,\mu{\rm as}} \right)^{3/2}.
\end{align}
For \sgra, a similar equation holds, accounting for the substantially higher compact flux density and somewhat larger photon ring, but the equation must also be multiplied by the diffractive scattering kernel \cite{Psaltis_2018,Johnson_2018}.

\begin{table}[ht]
\caption{{\bf Representative flux densities of primary BHEX targets.}
The total (compact) flux densities for \m87 and \sgra are based on interferometric observations of both targets \cite{Lu_2011,Bower_2015,EHTC_M87_IV}. The BHEX flux densities are based on values measured in GRMHD simulations and simple analytic arguments.} 
\label{tab:BHEX_Flux_Density}
\begin{center}       
\begin{tabular}{|l|l|l|} 
\hline
\rule[-1ex]{0pt}{3.5ex} & {\bf Total Flux Density} & {\bf BHEX Flux Density} \\
\rule[-1ex]{0pt}{3.5ex} {\bf \m87} &  &  \\
\rule[-1ex]{0pt}{3.5ex} \quad 80\,GHz & 0.6\,Jy & 40\,mJy\\
\rule[-1ex]{0pt}{3.5ex} \quad 240\,GHz & 0.5\,Jy & 10\,mJy \\
\rule[-1ex]{0pt}{3.5ex} \quad 320\,GHz & 0.35\,Jy & 5\,mJy \\
\hline
\rule[-1ex]{0pt}{3.5ex} {\bf \sgra} &  &  \\
\rule[-1ex]{0pt}{3.5ex} \quad 80\,GHz & 3\,Jy & 3\,mJy\\
\rule[-1ex]{0pt}{3.5ex} \quad 240\,GHz & 3.5\,Jy & 5\,mJy \\
\rule[-1ex]{0pt}{3.5ex} \quad 320\,GHz & 3.5\,Jy & 10\,mJy \\
\hline 
\rule[-1ex]{0pt}{3.5ex} $T_{\rm b}=10^{12}\,{\rm K}$ & 100\,mJy  & 88\,mJy \\
\rule[-1ex]{0pt}{3.5ex}  & 50\,mJy  & 47\,mJy \\
\rule[-1ex]{0pt}{3.5ex}  & 10\,mJy  & 9.9\,mJy \\
\hline 
\rule[-1ex]{0pt}{3.5ex} $T_{\rm b}=10^{11}\,{\rm K}$ & 100\,mJy  & 29\,mJy \\
\rule[-1ex]{0pt}{3.5ex}  & 50\,mJy  & 27\,mJy \\
\rule[-1ex]{0pt}{3.5ex}  & 10\,mJy  & 8.8\,mJy \\
\hline 
\rule[-1ex]{0pt}{3.5ex} $T_{\rm b}=10^{10}\,{\rm K}$ & 100\,mJy  & 0\,mJy \\
\rule[-1ex]{0pt}{3.5ex}  & 50\,mJy  & 0.1\,mJy \\
\rule[-1ex]{0pt}{3.5ex}  & 10\,mJy  & 2.9\,mJy \\
\hline 
\end{tabular}
\end{center}
\end{table}

\subsubsection{Requirements for Non-Photon-Ring Science and Calibration Targets}

For BHEX observations of marginally resolved LLAGN or knots in blazar jets, the flux density is limited by the maximum brightness temperature of incoherent synchrotron emission: $T_{\rm b} \lsim 10^{11}\,{\rm K}$ \cite{Kellermann_1969,Readhead_1994}.
Note that the observed brightness temperature can exceed this limit by an order of magnitude because of relativistic Doppler effects, as is commonly seen in blazar jets.
Since bright, marginally resolved sources are also ideal calibration targets for BHEX, this type of source is also useful to define mission requirements for calibrators.

For a Gaussian distribution of flux density, the central (peak) brightness temperature is $T_{\rm b,max} = \frac{2 \ln(2) c^2}{\pi \nu^2 k_{\rm B} \theta^2} F_0$.
Putting in characteristic values, 
\begin{align}
    T_{\rm b,max} = \left( 1.45 \times 10^{10}~{\rm K} \right) \times \left( \frac{\nu}{230~{\rm GHz}} \right)^{-2} \left( \frac{F_0}{1~{\rm Jy}} \right) \left( \frac{\theta}{40~\mu{\rm as}} \right)^{-2}.
\end{align}
The visibility domain response takes the form:
\begin{align}
\label{eq:V_gaussian}
    V(u) &= F_0 \exp\left( -\frac{\pi b^2 F_0}{2k_{\rm B} T_{\rm b}} \right).
\end{align}
Critically, the visibility amplitude depends only on the (projected) {\it physical} baseline length $b$ (rather than the dimensionless length $b/\lambda$).
On a fixed baseline $b$, Equation~\ref{eq:V_gaussian} is maximized when $F_0 = \frac{2 k_{\rm B} T}{\pi b^2}$.
Thus, under the assumption of a Gaussian emitting region, this type of source gives a peak correlated flux density that only depends on the brightness temperature and physical baseline length:
\begin{align}
    |V(b)| \leq (300\,{\rm mJy}) \times \left( \frac{T_{\rm b}}{10^{11}\,{\rm K}} \right) \left( \frac{b}{10^4\,{\rm km}} \right)^{-2}.
\end{align}
Hence, BHEX will primarily be sensitive to compact emitting regions with $T_{\rm b} \gsim 10^{11}\,{\rm K}$ and $F_0\lsim 100\,{\rm mJy}$ (see Table~\ref{tab:BHEX_Flux_Density}).
We note that this limit applies exclusively to marginally resolved emitting regions.
For heavily resolved regions, the flux density can be substantially higher than the Gaussian prediction from the overall source morphology (e.g., in the case of the photon ring) or substructure within the image. 

\section{Instrument and Technical Heritage}
\label{sec:Instrument}

BHEX builds upon the lessons of space-VLBI missions at centimeter wavelengths: TDRSS \cite{Levy_1986}, VSOP \cite{Hirabayashi_1998}, and RadioAstron \cite{Kardashev_2013} (for a brief review of space VLBI, see Refs.~[\citenum{TMS,Gurvits2020,Gurvits_2023}]).
The BHEX instrument is a radio telescope and (heterodyne) receiver system, similar to the corresponding ground systems at all EHT stations [for an overview of these systems, see Ref.~[\citenum{EHTC_M87_II}].  

In this section, we briefly summarize each of the six subsystems of the BHEX instrument: the antenna (Section~\ref{sec:Antenna}), the receiver (Section~\ref{sec:Receiver}), the cryocooler (Section~\ref{sec:Cryo}), the digital back end (Section~\ref{sec:DBE}), the frequency reference (Section~\ref{sec:freq_reference}), and the optical data transmission (Section~\ref{sec:lasercom}). Each of these subsystems leverages substantial space heritage and has been designed after considering a broad trade space; see Ref.~[\citenum{BHEX_Peretz_2024}]. A more detailed description of the BHEX instrument is given in Ref.~[\citenum{BHEX_Marrone_2024}].

\subsection{Antenna}
\label{sec:Antenna}

The BHEX antenna will have a rigid 3.5 m circular aperture. To minimize the mass, the antenna will employ a metallized carbon fiber reinforced plastic (CFRP) sandwich construction. This approach has been successfully used in past missions including Planck\cite{Tauber_2010}. To achieve our required sensitivity at the highest observing frequency ($320\,$GHz), a high surface accuracy with less than $40\,\mu{\rm m}$ rms is targeted. The current optical design incorporates a symmetric, dual reflector configuration that achieves an aperture illumination efficiency of $\eta_{\rm a} \approx 90\%$. Further exploration of design choices and parameter space and reflector shaping studies are in progress.  

For additional details on the space heritage, trade space, and design, see Refs.~[\citenum{BHEX_Sridharan_2024,BHEX_IEEE_Lehmensiek_2024}].

\subsection{Receiver}
\label{sec:Receiver}

BHEX includes two dual-polarization receivers together with an optical diplexer that allows them to observe simultaneously. The primary receiver will be double-side-band (DSB) operating over the frequency range 240-320\,GHz with a Superconductor-Insulator-Superconductor (SIS) mixer. SIS mixers are standard in ground submillimeter telescopes, were used in the Herschel mission \cite{Pilbratt_2010,Herschel_HIFI}, and are being actively developed for a variety of terrestrial and space applications\cite{Kojima_2018,BHEX_Tong_2024,BHEX_Akiyama_2024}. The secondary receiver will be single-side-band (SSB) operating over the frequency range 80-106\,GHz with a cryogenic low noise amplifier, which are commercially available.

On baselines from BHEX to ground stations, rapid phase variations are dominated by fluctuations in the troposphere and from errors in reference oscillators. Because all these contributions are non-dispersive delays, measurements at low frequencies can be used to stabilize the phase at high frequencies, substantially increasing the coherence time and, hence, the achievable sensitivity. This technique, known as frequency phase transfer, has been used routinely in centimeter observations \cite{Rioja_2015}, and is a major focus of ongoing EHT upgrades \cite{Rioja_2015,Issaoun_2023,Rioja_2023,Jiang_2023}. The receivers of BHEX are designed to enable this technique, transferring phase measurements from the secondary band to derive phase corrections for the primary band. This design both improves the sensitivity of BHEX and mitigates the requirements for the reference frequency. 

For additional details on the space heritage, trade space, and design of the BHEX receivers, see Ref.~[\citenum{BHEX_Tong_2024}].

\subsection{Cryocooler}
\label{sec:Cryo}

The BHEX receiver suite requires cryogenic cooling to achieve the required receiver noise temperatures, close to the quantum noise limit $T_{\rm Q} = h\nu/k$ (see Section~\ref{sec:Receiver}). The SIS mixers of the primary receiver, in particular, must be cooled to 4.5\,K (approximately half the critical temperature of the SIS junction), while the low noise amplifier of the secondary receiver performs better when cooled to 20\,K. It is worth noting that BHEX only requires a modest heat lift to cool the receiver suite and does not need to cool the dish.  BHEX aims to leverage existing cryogenic technology, where the majority of the key cryogenic components within these technologies boast established spaceflight heritage. Recent missions (e.g., SMILES/JEM, ASTRO-H, XRISM) include closed-cycle 4\,K cryocoolers that are a good match to the BHEX design and requirements.

For additional details on the space heritage, trade space, and design of the BHEX cryocooler, see Ref.~[\citenum{BHEX_Rana_2024}].

\subsection{Block Downconverter and Digital Back End}
\label{sec:DBE}

Following the receiver, the digital stage of BHEX requires a block downconverter (BDC) and a digital back end that can sample and digitize a combined 32\,GHz of analog bandwidth. Because the sampled signal is a zero-mean Gaussian random field (corresponding to band-limited noise from both the source and thermal background noise), no additional data compression can be applied before downlink, aggregation with the data from other participating observatories, and correlation \cite{TMS}.

The BHEX digital back end is informed by state-of-the art EHT designs, which must operate in harsh terrestrial environments (e.g., at the South Pole or at an altitude of 5000 meters for the ALMA Observatory). These systems use commercial wideband samplers and Field Programmable Gate Arrays (FPGAs) to enable a relatively low-power and low-cost design. The primary difference for the BHEX system is the requirements for space-grade components, which are also commercially available. Another difference is that the BHEX DBE will use a 1-bit quantization scheme, giving a quantization efficiency of $\eta_{\rm Q}=0.750$ on baselines from BHEX to ground sites with 2-bit quantization. 


For additional details on the space heritage, trade space, and design of the BHEX DBE, see Ref.~[\citenum{BHEX_Srinivasan_2024}].

\subsection{Frequency Reference}
\label{sec:freq_reference}

VLBI relies on a stable frequency reference to allow coherent integration of the correlated signal between sites in order to find ``fringes.'' With a sufficiently stable reference, the integration time is limited by fluctuations in optical path length through the troposphere, with typical coherence times of ${\sim}10$ second at wavelengths of $\lambda \approx 1.3\,{\rm mm}$ for EHT sites. With sufficient signal-to-noise on short integrations, residual fluctuations can be corrected in post-processing (a procedure sometimes called ``ad-hoc phase correction'') \cite{EHTC_M87_III,Blackburn_2019}.

With the use of multi-band observations, using a secondary band for phase corrections, BHEX can use an ultra stable oscillator (USO) based on a temperature-controlled quartz crystal. For example, an exceptionally quiet USO \cite{Accubeat_press, NIST_disclaimer} is currently serving as a reference for the European Space Agency's (ESA) Jupiter Icy Moons Explorer (JUICE) mission\cite{juice}. We are also currently evaluating alternative choices of frequency reference that use the stabilization of a continuous wave (CW) laser to an optical atomic or molecular transition.  By referencing the output of an optical frequency comb to stabilized CW laser, the coherent optical pulse train output from the comb serves as the clock output and a microwave reference frequency can be generated which preserves the fractional frequency instability of the CW laser. Some of these optical atomic clocks can be quite simple, with performance that exceeds that of an active hydrogen maser over the timescales of interest \cite{roslund_optical_2024}.

For additional details on the space heritage, trade space, and design of the BHEX DBE, see Refs.~[\citenum{BHEX_Peretz_2024,BHEX_Marrone_2024}].

\subsection{Optical Data Transmission}
\label{sec:lasercom}

Space optical communications has long been recognized as capable of supporting higher data rates with smaller apertures and lower powers than the incumbent radio-frequency communications methods widely used today.  
Over the past several decades, these capabilities have been demonstrated over a variety of orbits and mission use cases such as lunar missions for science \cite{Boroson_2014}, geosynchronous relays \cite{Caplan_2010,Spellmeyer_2014, Edwards_2018}, and low-earth orbit and lunar missions supporting human exploration \cite{Khatri_ISS_2023,Khatri_lunar_2023}.  One unique aspect of space optical communications addressed by all missions is the sensitivity of optical frequencies to atmospheric scintillation and fading, requiring atmospheric mitigation techniques to provide error-free data delivery. To support high-fidelity downlink of data captured at up to 64\,Gb/s, BHEX requires a 100 Gb/s optical downlink system.

Recently, the TeraByte Infrared Delivery (TBIRD) mission, a small cubesat in low-earth orbit (LEO), demonstrated transmission up to 200\,Gb/s from space to ground \cite{Riesing_2023,Schieler_2023}. We have already explored adaptations to TBIRD that would meet requirements that are more demanding than the present BHEX design; see Ref.~[\citenum{Wang_2023}]. Such a mission can be achieved with technologies which have already been demonstrated today. 

For additional details on the space heritage, trade space, and design of the BHEX laser communications downlink design, see Ref.~[\citenum{BHEX_Wang_2024}].

\section{Concept of Operations}
\label{sec:ConOps}

The BHEX science goals hinge on building sensitivity from high recording bandwidths and optimal atmospheric conditions at the ground VLBI stations co-observing with the satellite. Weather considerations also need to be taken into account when developing the downlink network receiving the signal from the satellite. The high bandwidth requirements for BHEX necessitate a downlink infrastructure with optical communications, which have demonstrated higher bandwidth rates than radio-frequency methods commonly used in past space-VLBI experiments. The operations concept for BHEX essentially involves a three-part hybrid observatory: satellite operations of the space-based component, coordinated operations of the ground-based VLBI network, and coordinated operations of the ground-based downlink terminals.

\begin{figure}[t]
    \centering
    \includegraphics[width=\textwidth]{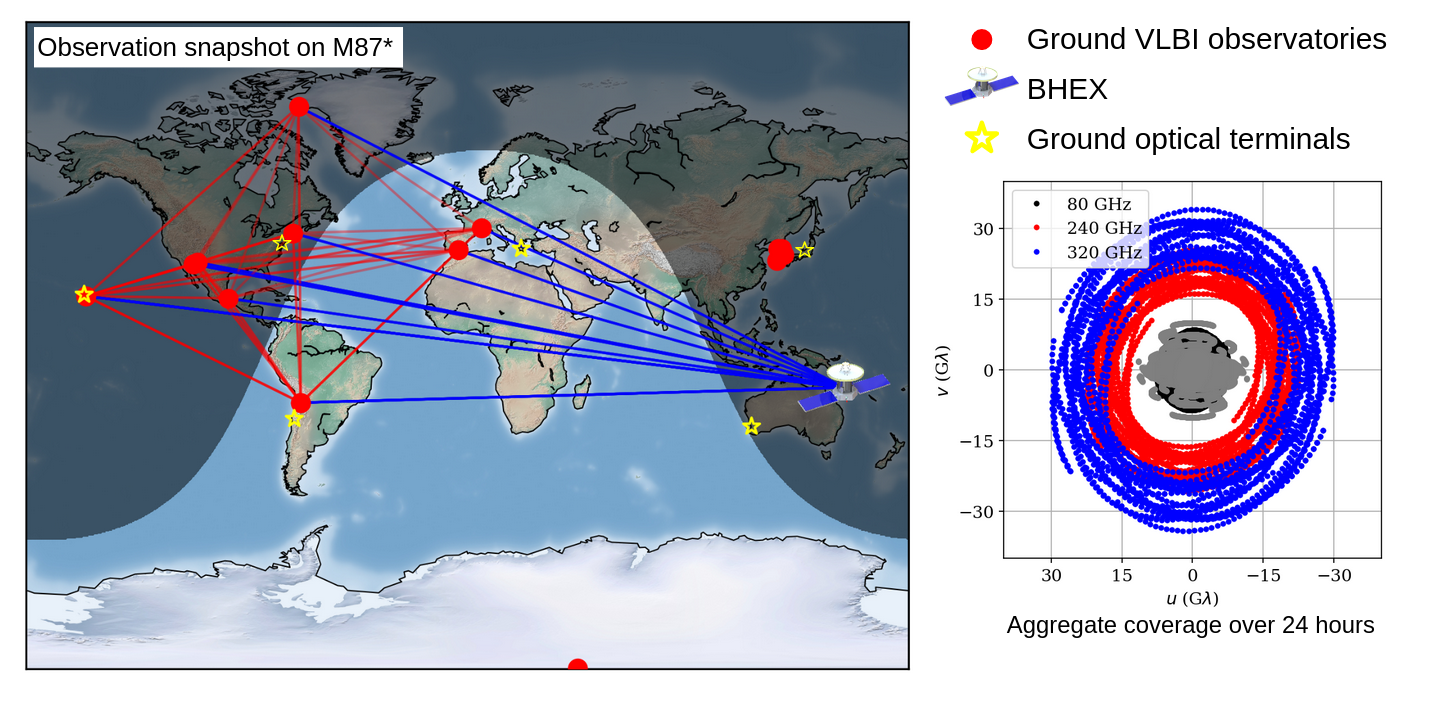}
    \caption{(left) The BHEX concept of operations for \m87. Red and blue lines represent co-observing of the target for ground-only and space-ground telescope pairs, respectively. (right) The geometrical $(u,v)$ coverage on \m87 over 24 hours, where BHEX baselines at 80, 240, and 320\,GHz are shown in black, red, and blue, respectively. Ground-ground baselines are shown in grey.}
\label{fig:SciOps_m87} 
\end{figure}

\begin{figure}[t]
    \centering
    \includegraphics[width=\textwidth]{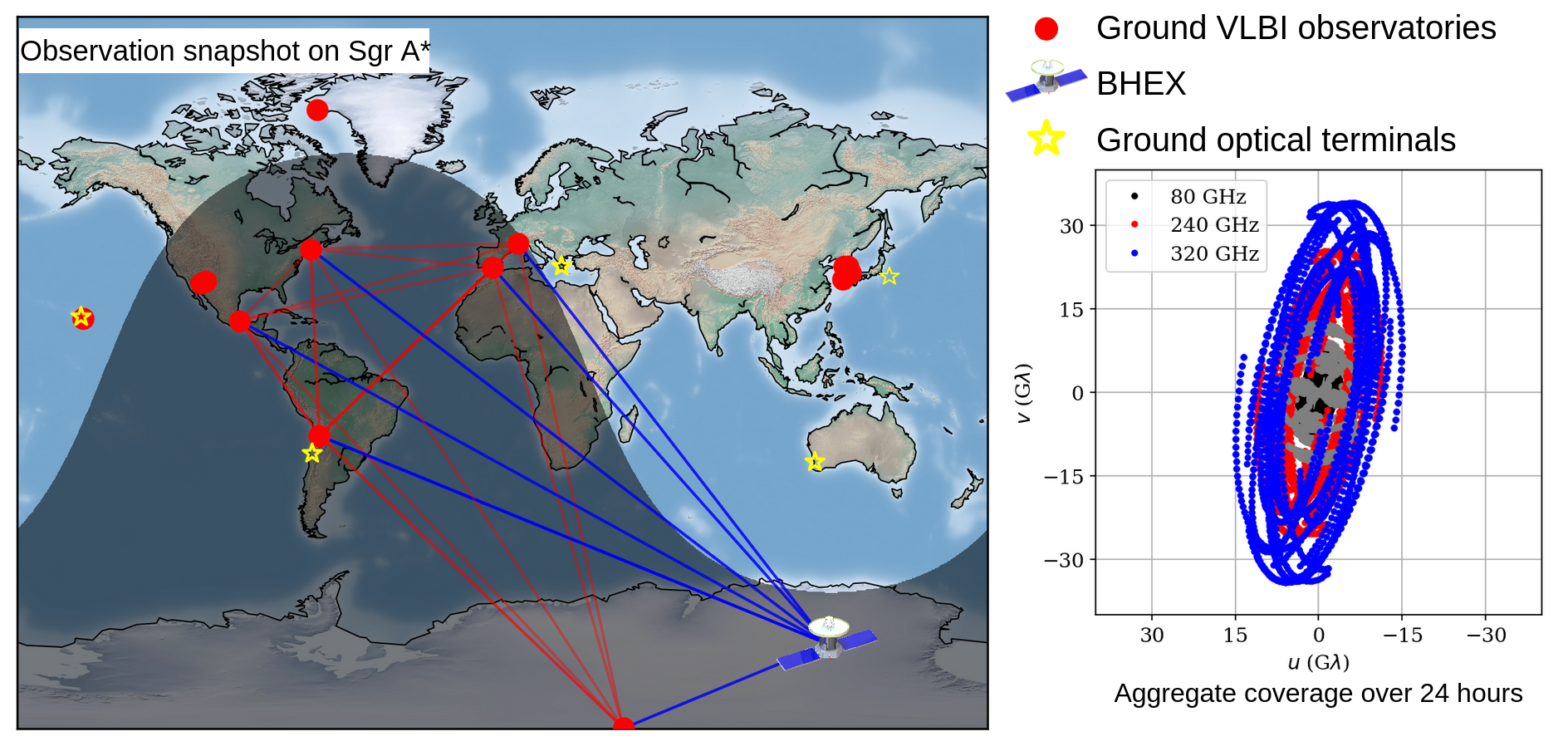}
    \caption{Same as Figure~\ref{fig:SciOps_m87}, but for \sgra. The orbit is selected to provide coverage that is elongated along the minor axis of the scattering kernel, oriented approximately $8^\circ$ west of north.}
\label{fig:SciOps_sgra} 
\end{figure}

BHEX will operate for a nominal two-year mission in a circular, semi-synchronous, near-polar orbit (\(\sim\)20,200~km altitude), optimized for multi-directional sampling
of the photon ring signal in \m87. Repeated observations are key to aggregate sufficient data to make images of the photon ring. This will provide the required baseline coverage and thus angular resolution of the primary targets, \m87 and \sgra. Observations of sources will be conducted at the times of year when the angle between the antenna boresight and the Sun is greater than 90 degrees to minimise thermal distortion of the antenna surface due to solar radiation. This requirement, and the favourable times of year for ground station observations, results in observation campaigns of the two main sources during the following months: January - March for \m87 and June - August for \sgra, respectively. Photon ring campaigns will be interspersed with survey science on demographics of black holes in low accretion states and relativistic jet studies of a range of sources.

Radio emission from distant black holes will be observed by the BHEX and telescopes on the ground simultaneously, including those that currently form the EHT. BHEX will leverage established (sub)millimeter observatories in the EHT to carry out its science goals. Large sensitive apertures (such as ALMA, the LMT, the MIT Haystack 37-m, the IRAM 30-m, and NOEMA telescopes, and the SMA) will provide detections to BHEX and anchor the satellite to the smaller ground dishes filling our coverage for imaging, see Figures~\ref{fig:SciOps_m87} and \ref{fig:SciOps_sgra} for \m87 and \sgra, respectively. Partnerships with these facilities are already established within the EHT collaboration, and a path for time allocation is being pursued for BHEX. The observed radio waves will be recorded at the ground-based telescopes and will be computationally combined with the observational data transmitted to the ground from BHEX to generate the highest resolution images of black holes. BHEX will perform real time downlink of data during observations utilising a network of optical ground terminals, distributed across the globe (see section~\ref{sec:lasercom}). The semi-synchronous orbit selection is favourable for this mode of operation as it generates a constant ground track which, alongside a careful selection of ground terminal sites, ensures that a real time downlink can be maintained at all times.

We plan to utilize existing EHT correlation infrastructure, such as the correlator at MIT Haystack Observatory and the Cannon Cluster at the Massachusetts Green High Performance Computing Center (MGHPCC). The correlated data will then be processed using the EHT reference pipeline, and both raw, correlated data products and calibrated, processed data products will be released to a public archive.

\begin{table}[t]
\caption{{\bf BHEX Mission Characteristics.} 
} 
\label{tab:BHEX_Mission}
\begin{center}       
\begin{tabular}{|l|l|l|} 
\hline
\rule[-1ex]{0pt}{3.5ex} {\bf Antenna} & \\
\rule[-1ex]{0pt}{3.5ex} \quad Diameter & $3.5$-m\\
\rule[-1ex]{0pt}{3.5ex} \quad Surface RMS & 40\,$\mu$m\\
\hline
\rule[-1ex]{0pt}{3.5ex} {\bf Primary Receiver} (SIS; 4.5\,K) & \\
\rule[-1ex]{0pt}{3.5ex} \quad Frequency Range & 240-320\,GHz \\
\rule[-1ex]{0pt}{3.5ex} \quad IF & 4-12\,GHz \\
\rule[-1ex]{0pt}{3.5ex} \quad Receiver noise temperature (DSB) & 23-30 K\\
\rule[-1ex]{0pt}{3.5ex} \quad SEFD at 240\,GHz & 16{,}400~Jy \\
\rule[-1ex]{0pt}{3.5ex} \quad SEFD at 320\,GHz & 23{,}600~Jy  \\
\hline
\rule[-1ex]{0pt}{3.5ex} {\bf Secondary Receiver} (HEMT; 20\,K) & \\
\rule[-1ex]{0pt}{3.5ex} \quad Frequency Range & 80-106\,GHz \\
\rule[-1ex]{0pt}{3.5ex} \quad IF & 4-12\,GHz \\
\rule[-1ex]{0pt}{3.5ex} \quad Receiver noise temperature (SSB) & 45 K\\
\rule[-1ex]{0pt}{3.5ex} \quad SEFD at 80\,GHz & 14{,}700~Jy \\
\hline
\rule[-1ex]{0pt}{3.5ex}  {\bf Maximum Data Rate} & 64 Gb/s (both receivers; dual-pol; 1-bit quantization)\\
\hline
\rule[-1ex]{0pt}{3.5ex} {\bf Baseline Sensitivity} & (RMS Thermal Noise for 8\,GHz Bandwidth) \\
\rule[-1ex]{0pt}{3.5ex} \quad BHEX-ALMA, 10-minute integration  & 1\,mJy \\
\rule[-1ex]{0pt}{3.5ex} \quad BHEX-ALMA, 10-second integration  & 7\,mJy \\
\rule[-1ex]{0pt}{3.5ex} \quad BHEX-LMT, 10-minute integration  & 2\,mJy \\
\rule[-1ex]{0pt}{3.5ex} \quad BHEX-LMT, 10-second integration  & 15\,mJy \\
\hline
\rule[-1ex]{0pt}{3.5ex} {\bf Orbit} & Semi-synchronous (12-hr; 20{,}200\,km altitude)\\ 
\rule[-1ex]{0pt}{3.5ex} &  Circular\\
\rule[-1ex]{0pt}{3.5ex} &  ${\geq}78^\circ$ inclination  \\
\hline
\rule[-1ex]{0pt}{3.5ex} {\bf Finest Fringe Spacing} ($\lambda/D$) &  \\
\rule[-1ex]{0pt}{3.5ex} \quad 80\,GHz  & $24\,\mu{\rm as}$\\
\rule[-1ex]{0pt}{3.5ex} \quad 240\,GHz & $8\,\mu{\rm as}$\\
\rule[-1ex]{0pt}{3.5ex} \quad 320\,GHz & $6\,\mu{\rm as}$\\
\hline
\rule[-1ex]{0pt}{3.5ex} {\bf Maximum Baseline Length} & $32{,}900\,{\rm km}$\\
\rule[-1ex]{0pt}{3.5ex} \quad 80\,GHz  & $9\,{\rm G}\lambda$\\
\rule[-1ex]{0pt}{3.5ex} \quad 240\,GHz & $26\,{\rm G}\lambda$\\
\rule[-1ex]{0pt}{3.5ex} \quad 320\,GHz & $35\,{\rm G}\lambda$\\
\hline
\rule[-1ex]{0pt}{3.5ex} {\bf Mission Lifetime} & $2\,{\rm yr}$ \\
\hline 
\end{tabular}
\end{center}
\end{table}

\section{Summary}
\label{sec:Summary}

The BHEX mission concept, summarized in Table~\ref{tab:BHEX_Mission}, arises from a coincidence of transformational theoretical discoveries, an explosion of community interest in black holes, and technical breakthroughs that enable submillimeter space VLBI.
BHEX will produce images that bring us to an edge of the universe, revealing whether spinning black holes are feeding energy back into the universe by powering relativistic jets of emission.
And it will reveal a population of unseen black holes, showing how they grow over cosmic timescales. 

BHEX is enabled by pioneering work to develop successful space-VLBI missions at centimeter wavelengths and emerging space technologies such as laser communications.
It leverages billions of dollars of existing ground infrastructure that has already been combined to deliver the remarkable successes of the EHT.
With these advances, we are now at a moment when a critical new science goal---imaging a black hole's photon ring---is within reach.
Guided by this common vision, the BHEX team will submit a Small Explorers mission proposal in 2025 to launch BHEX within the next decade and turn this extraordinary opportunity into a reality.

\appendix

\clearpage

\acknowledgments 

Technical and concept studies for BHEX have been supported by the Smithsonian Astrophysical Observatory, by the Internal Research and Development (IRAD) program at NASA Goddard Space Flight Center, by the University of Arizona, and by the ULVAC-Hayashi Seed Fund from the MIT-Japan Program at MIT International Science and Technology Initiatives (MISTI). We acknowledge financial support from the Brinson Foundation, the Gordon and Betty Moore Foundation (GBMF-10423), the National Science Foundation (AST-2307887, AST-2307888, AST-2107681, AST-1935980, and AST-2034306), the Simons Foundation (MP-SCMPS-00001470), 
the European Research Council (the European Union’s Horizon 2020 Research and Innovation Programme, grant No 101018682), and the Israel Science Foundation (grant \#2047/23).
This project/publication is funded in part by the Gordon and Betty Moore Foundation (Grant \#8273.01). It was also made possible through the support of a grant from the John Templeton Foundation (Grant \#62286).  The opinions expressed in this publication are those of the author(s) and do not necessarily reflect the views of these Foundations. BHEX is funded in part by generous support from Mr. Michael Tuteur and Amy Tuteur, MD. BHEX is supported by initial funding from Fred Ehrsam.

\bibliography{report,bhexspiepapers} 

\begin{thebibliography}{100}
\bibitem{NIST_disclaimer}
``Certain equipment or instruments are identified in this paper in order to specify the experimental procedure adequately. such identification is not intended to imply recommendation or endorsement of any product by nist, nor is it intended to imply that the equipment identified are necessarily the best available for the purpose..''
\end{thebibliography}
\bibliographystyle{spiebib} 

\end{document}